%% file: sample-acmlarge.tex
\documentclass[acmlarge]{acmart}

\usepackage{booktabs} 
\usepackage{subcaption}

\usepackage[ruled]{algorithm2e} 

\SetAlFnt{\small}
\SetAlCapFnt{\small}
\SetAlCapNameFnt{\small}
\SetAlCapHSkip{0pt}
\IncMargin{-\parindent}

\acmJournal{IMWUT}
\acmVolume{0}
\acmNumber{0}
\acmArticle{0}
\acmYear{2018}
\acmMonth{2}
\acmArticleSeq{0}

\setcopyright{usgovmixed}

\acmDOI{0000001.0000001}

\begin{document}
\title{Predicting Clinical Deterioration of Outpatients Using Multimodal Data Collected by Wearables} 

\author{Dingwen Li}
\affiliation{%
  \institution{Washington University in St. Louis}
  \streetaddress{1 Brookings Dr}
  \city{St. Louis}
  \state{MO}
  \postcode{63130}
  \country{USA}}
\author{Jay Vaidya}
\affiliation{%
  \institution{Washington University in St. Louis}
  \streetaddress{1 Brookings Dr}
  \city{St. Louis}
  \state{MO}
  \postcode{63130}
  \country{USA}}
\author{Michael Wang}
\affiliation{%
  \institution{Washington University in St. Louis}
  \streetaddress{1 Brookings Dr}
  \city{St. Louis}
  \state{MO}
  \postcode{63130}
  \country{USA}}
\author{Ben Bush}
\affiliation{%
  \institution{Washington University in St. Louis}
  \streetaddress{1 Brookings Dr}
  \city{St. Louis}
  \state{MO}
  \postcode{63130}
  \country{USA}}
\author{Chenyang Lu}
\affiliation{%
  \institution{Washington University in St. Louis}
  \streetaddress{1 Brookings Dr}
  \city{St. Louis}
  \state{MO}
  \postcode{63130}
  \country{USA}}
\author{Marin Kollef}
\affiliation{%
  \institution{Washington University in St. Louis}
  \streetaddress{1 Brookings Dr}
  \city{St. Louis}
  \state{MO}
  \postcode{63130}
  \country{USA}}
\author{Thomas Bailey}
\affiliation{%
  \institution{Washington University in St. Louis}
  \streetaddress{1 Brookings Dr}
  \city{St. Louis}
  \state{MO}
  \postcode{63130}
  \country{USA}}

\begin{abstract}
Hospital readmission rate is high for heart failure patients. Early detection of deterioration will help doctors prevent readmissions, thus reducing health care cost and providing patients with just-in-time intervention. Wearable devices (e.g., wristbands and smart watches) provide a convenient technology for continuous outpatient monitoring. In the paper, we explore the feasibility of monitoring outpatients using Fitbit Charge HR wristbands and the potential of machine learning models to predicting clinical deterioration (readmissions and death) among outpatients discharged from the hospital. We developed and piloted a data collection system in a clinical study which involved 25 heart failure patients recently discharged from a hospital. The results from the clinical study demonstrated the feasibility of continuously monitoring outpatients using wristbands. We observed high levels of patient compliance in wearing the wristbands regularly and satisfactory yield, latency and reliability of data collection from the wristbands to a cloud-based database. Finally, we explored a set of machine learning models to predict deterioration based on the Fitbit data. Through 5-fold cross validation, K nearest neighbor achieved the highest accuracy of 0.8800 for identifying patients at risk of deterioration using the health data from the beginning of the monitoring. Machine learning models based on multimodal data (step, sleep and heart rate) significantly outperformed the traditional clinical approach based on LACE index. Moreover, our proposed weighted samples one class SVM model can reach high accuracy (0.9635) for predicting the deterioration happening in the future using data collected by a sliding window, which indicates the potential for allowing timely intervention.
\end{abstract}

%
%

\ccsdesc[500]{Human-centered computing~Ubiquitous and mobile computing systems and tools}

%
%

\keywords{Mobile sensing, wearables, hospital readmission, clinical deterioration, machine learning, heart failure}


\maketitle

\renewcommand{\shortauthors}{}

\input{samplebody-journals}

\end{document}

%% file: samplebody-journals.tex
\section{Introduction}

Hospital readmissions occur often and are difficult to predict \cite{marks13}. Many hospitals have dedicated resources to identify patients at risk for readmission, and to prevent such readmissions \cite{marks13}. Heart failure is the most common principle hospital discharge diagnosis among Medicare beneficiaries and is among the most expensive conditions billed to Medicare \cite{andrews07}. Readmission rates following discharge for heart failure are high, with approximately $25\%$ of patients being readmitted within 30 days. However, only about $35\%$ of these patients are readmitted for heart failure \cite{Dharmarajan13}, and efforts to prevent these readmissions, most of which have focused on the condition causing the index admission, have met with variable and incomplete success \cite{Ziaeian15}. It is difficult to accurately diagnose heart failure in time \cite{MaDonald01}. In the medical field, clinicians use standardized LACE index (calculated based on length of stay, acuity of admission, co-morbidity index and number of emergency department visit) to evaluate the risk of readmission or death after discharge. The problems of frequent readmission and hard-to-detect deterioration in patients with heart failure, along with the health care costs associated with these readmissions, suggests the need for innovative, global monitoring strategies that call attention to patients who are in need of attention to prevent clinical deterioration and consequent need for emergency department visits or hospital readmission. 

Wearable devices have become accessible for general population. While activity, heart rate and sleep quality sensors (wristbands and smart watches) are increasingly popular among healthy individuals, few studies have assessed their potential role in health care \cite{Appelboom15, Vooijs14}. The more common accessibility of wearable devices leads an evolution in medical research, which helps clinicians collect everyday health data in a cheap and friendly way. Studies showed Fitbit worked well in activity related monitoring and data collection \cite{cadmus15}. Wearable devices can achieve pretty high accuracy in steps measurement. An earlier study found that Fitbit Flex wristbands had an accuracy of $0.996$ when measuring straight indoor walking \cite{Kaewkannate2016}. Wearable devices also show their accuracy in measuring everyday heart rate with errors ranging from $1.8\%$ to $5.5\%$ \cite{Shcherbina17}. Due to their attractive designs, easy to use characteristics and accuracy, wearable fitness devices are promising devices for monitoring outpatients.

In this paper we explore the potential of wearables to monitor clinical deterioration among outpatients. Specifically, we aim to predict \textit{clinical deterioration} defined as a composite outcome of either readmission or death among patients discharged from a hospital. We first developed a cloud-based database system to collect multimodal data from outpatients using the Fitbit Charge HR wristband. The system can passively collect everyday health data including step counts, heart rates, and sleep duration and quality. We then conducted a clinical study to monitor 25 heart failure patients recently discharged from a major research hospital in United States. We assessed the feasibility of collecting multimodal data from outpatients using wearables by analyzing the data yield, reliability, and patient compliance of our data collection system based on wristbands. Finally, we demonstrate the potential of applying machine learning to predict clinical deterioration among outpatients as either predicting deterioration ahead of time with recent data or identifying patient at risk of deterioration with data from beginning of monitoring. 

The main findings of our study are two-fold. First, our experience in the clinical study demonstrated the feasibility of collecting multimodal data from outpatients using wearables. In the study, our system collected more than $80\%$ of per-minute step data from $84\%$ of the patients. While the yield of heart rate data is lower, the median gap in hear rate data was only 4 minutes. The results suggests a high level of compliance of patients in wearing the wristband regularly. Furthermore, the cloud-based database collected $73\%$ of the data within an hour, which allows timely intervention. 

Second, the superior performance of weighed samples one class SVM (with accuracy of 0.9635) demonstrates the feasibility of predicting clinical deterioration ahead of time. In the feasibility study of identifying patients at risk of deterioration, the performance evaluation of a set of predictive models show that the K nearest neighbor model achieved significantly higher accuracy ($0.8800$) than the traditional method based on LACE index ($0.7826$). Furthermore, combining multiple modalities (step, heart rate and sleep) effectively improved predictive accuracy when compared to models based on a single modality, which suggests the significance of incorporating multiple modalities for training predictive models.

\section{Related Work}

There has been significant interest in predicting clinical deterioration. Based on the sources of data used to perform the prediction, studies in this area can be categorized as prediction using clinical data collected within the hospital and that based on data collected by wearable devices. 

Clinicians traditionally relied on scores derived from inpatient data ~\cite{Walraven551,Wang2014,Robinson2017,Low2015,Cotter2012,Kansagara11,Donzé2016} or patient mobility ~\cite{Fisher2013,TAKAHASHI2015286}. In particular, LACE index has been widely adopted to predict readmission. The LACE index is manually calculated at the time of discharge to predict the risk of readmission of both medical and surgical patients after hospital discharge\cite{Wang2014}. However, LACE index has variable performance for different use cases. Robinson and Gudali \cite{Robinson2017} showed that LACE index had fair discrimination in a study of 5,800 patients in Singapore \cite{Low2015} and poor discrimination in a study done on about 500 patients in UK with an average age of 85 years \cite{Cotter2012}. Wang \cite{Wang2014} claimed that LACE index might not accurately predict 30-day readmission of congestive heart failure patients discharged from hospital. LACE-rt, as a real-time version of LACE index, was invented to predict readmission using the length of stay during the previous acute care admission \cite{ElMorr2017}. However, LACE-rt underestimates the readmission rates by not taking early death into account \cite{ElMorr2017}. 

As predicting hospital readmission via LACE index has variable performance, recent literature exploited learning-based approaches to train predictive models using clinical data. Logistic regression is a commonly used machine learning model to predict hospital readmission \cite{He2014,Sushmita2016,Low2015}. \cite{Sushmita2016,Low2015} demonstrated that logistic regression outperformed LACE index in predicting 30-day readmission, while several other studies \cite{Sushmita2016, Vedomske13,ZHENG20157110} showed that Random forest was also an accurate model in predicting hospital readmission. These learning-based approaches relied on medical records and vital signs collected while patients were \textit{in} the hospital \cite{Hosseinzadeh2013,Wang17,SHAMEER2016,Wang17,Zolfaghar13}. To date, there is scarce literature on predicting readmission using outpatient data.

As wearables can passively collect outpatient data in a continuous fashion, recent studies sought to employ wearables to predict readmission. Abdulmajeed et al. studied the feasibility of predicting the readmission of heart failure patients via wearables \cite{Abdulmajeed16}. Bae at al. \cite{Bae2016} demonstrated the feasibility of predicting readmission of postsurgical cancer via sedentary behavior data collected by Fitbit. However, both of the studies utilized features from a \textit{single} modality, i.e., step. As wearables incorporate more sensing modalities, it is important to explore the potential of \textit{multimodal} data to improve model accuracy. Our study combined features derived from heart rate and sleep data in addition to step data. Our comparative evaluation demonstrated the advantage of exploiting multimodal data for predicting clinical deterioration. Moreover, while previous studies using wearables focused on predictive models only, we further provided in-depth analysis of patient compliance and the performance of data collection from heart failure patients after hospital discharge. By demonstrating the feasibility of continuous data collection from outpatients using wearables, the findings from our clinical study have broad implications on future clinical studies that use wearables to monitor outpatients.


\begin{figure}
		\center
		\includegraphics[scale=0.3]{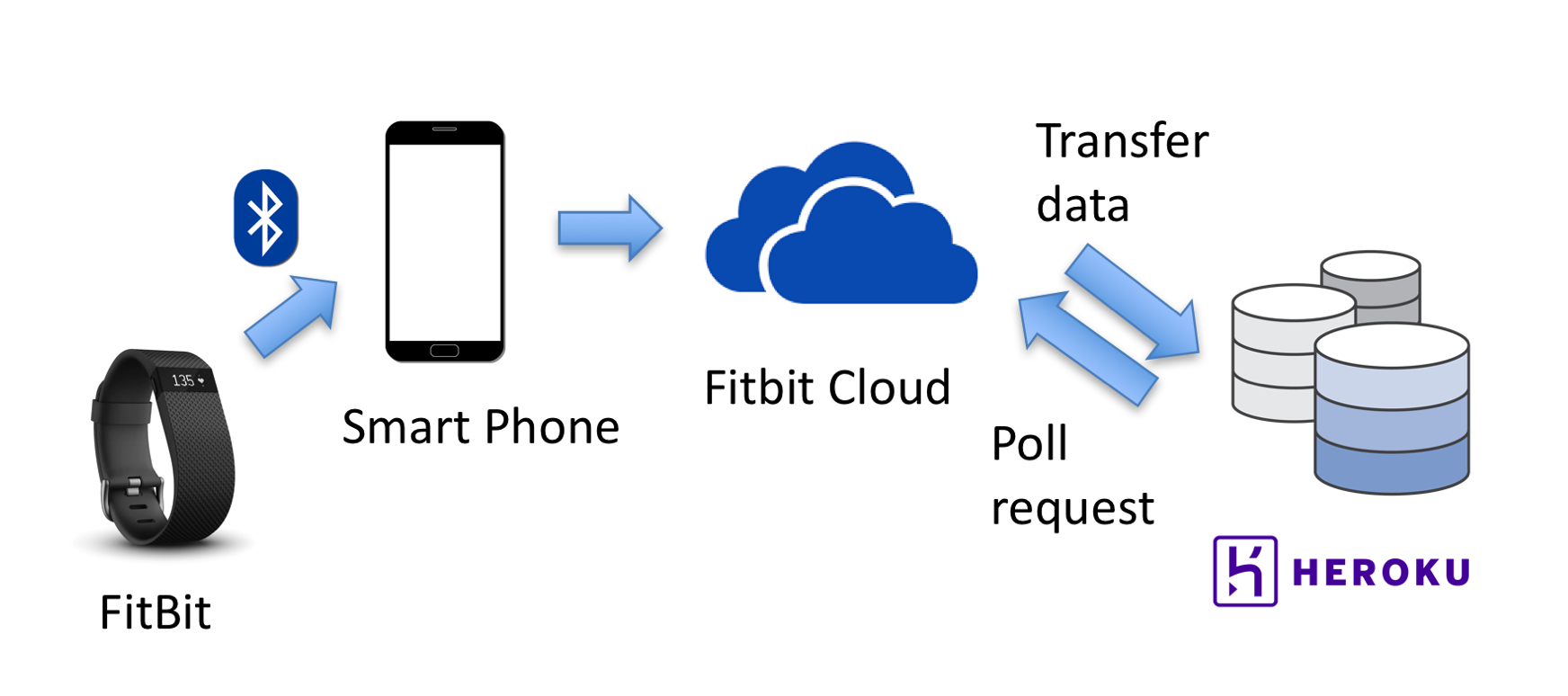}
        \caption{System Overview}\label{fig:system1}
\end{figure}

\section{System Design}

We have developed HIPAA-compliant software to collect multimodal data from Fitbit wristbands and store them a cloud-based database. The system architecture is shown in Figure \ref{fig:system1}. Each patient wears a Fitbit Charge HR wristband. The wristband collects multimodal data including per-minute heart rate and step count, the duration and quality of each sleep episode, and the remaining battery life of the wristband. The wristband synchronizes with the Fitbit App on an Internet connected device, in our case a smartphone, through Bluetooth. The smartphone then send the collected data to the Fitbit cloud where all user data is stored indefinitely. If the wristband is unable to synchronize with a smartphone and push the data to the Fitbit cloud, it will store intraday data locally on the device for up to 7 days. Therefore, as long as there is a synchronization between the Fitbit and smartphone within 7 days, there should be no data loss. Afterwards, it reduces to storing daily summary data for the next 30 days. The data communication from the wristband to the Fitbit cloud is managed by software provided by Fitbit. Once data has been stored in the Fitbit cloud, our own cloud-based Heroku server retrieves the data of our participants through Fitbit's developer API and stores it in our PostgreSQL database. If the data indicates that there may be issues with patient compliance, then our server uses Google's API and a Python Excel package to generate an Excel file containing the anonymized patient's data and sends an alert through Gmail to the nurses who are in charge of the study. The nurse reviews patient's data and contacts the patient if necessary. 

The end-to-end latency of data collection from the wristband to our database is influenced by the frequency at which data flows through the system. When the Fitbit App is setup on the patient's smartphone, the setting to synchronize the Fitbit Charge HR with the smartphone every 15 minutes is enabled. The two devices synchronize over Bluetooth so in order for data to be successfully transmitted, they must be within range of one another. Once the smartphone has received the data from the Fitbit, it immediately attempts to send this information to the Fitbit cloud over the Internet. From the Fitbit cloud, our application requests intraday heart rate and step data as well as accessory data (battery life at last sync and time of last sync) every 10 minutes for all patients. On a daily basis, it requests intraday and sleep summary data, and runs a checking routine to determine if patients are properly complying with the study protocol. If the system detects any anomalies, then it will send an email to the nurses and inform them to contact the patient. We used the number of total heart rate data points in the previous day as a proxy for what percentage of the day the patient was wearing the device properly. We decided on using heart rate data because it was the most granular with multiple samplings per minute. If either the number of heart rate data points was below 5400 per day or the last known battery level of the patient's device was below medium we alerted the nurses. 

\section{Performance of Data Collection}

In order to assess the feasibility of using wearables to monitor the health conditions of outpatients, it is important to evaluate the patients' compliance  rate and the yield, reliability and latency of the data collection process. 

\subsection{Yield}

The \textit{yield} is defined as the fraction of the expected samples that are successfully collected and stored in our database.  The yield is calculated separately for heart rate, step and sleep data. The sampling rate of heart rate and sleep is one sample per minute. For example, for a duration of 1 hour we are expecting 60 samples of heart rate and step count, respectively. To measure the yield of sleep data, we choose the sampling rate to be one sample per day. In that case, the sleep yield is defined as the fraction of days in which we collected sleep data.

\begin{figure} [h!]
\begin{subfigure}[h]{0.48\linewidth}
		\includegraphics[width=\linewidth]{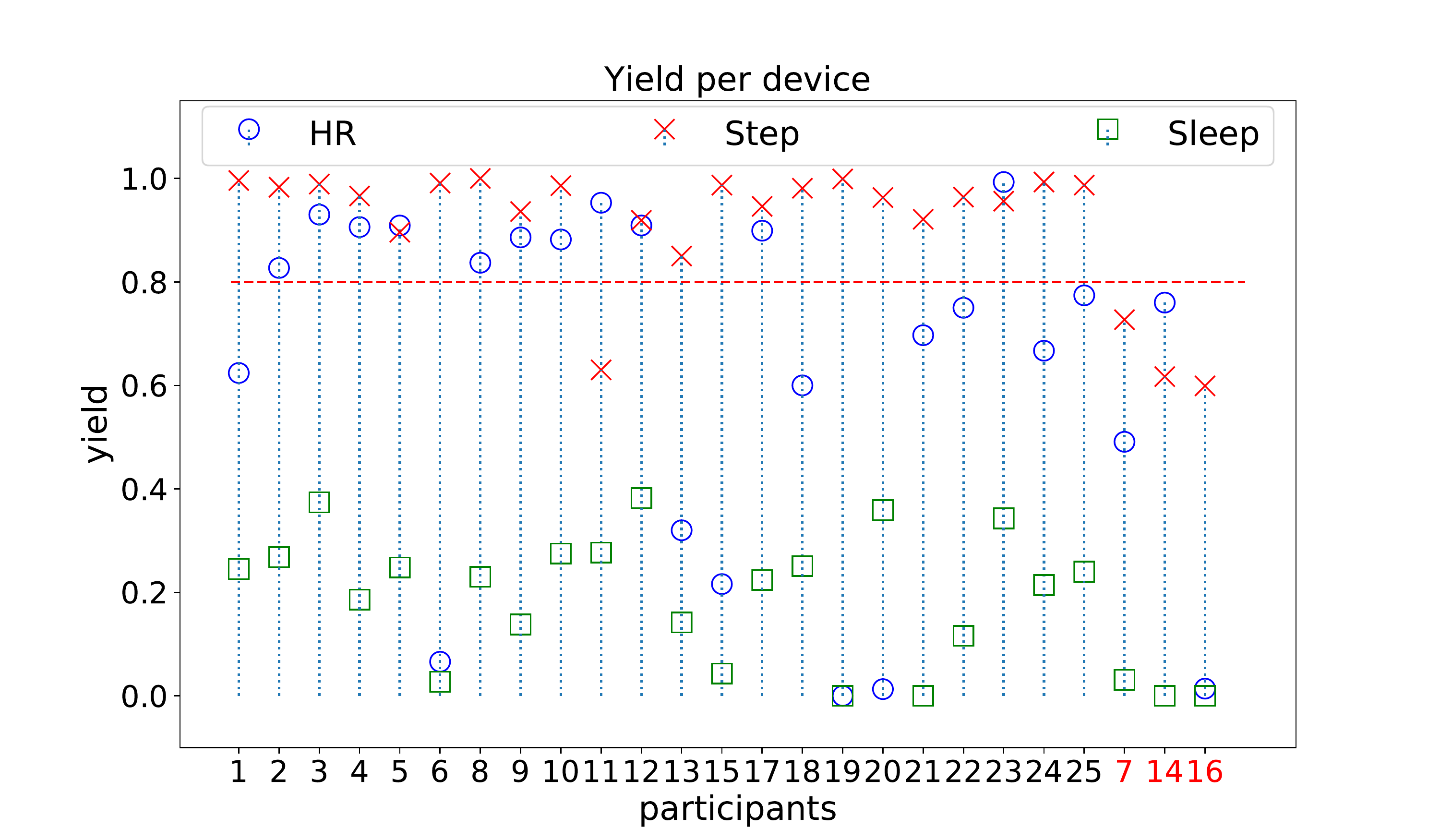}
\caption{}\label{fig:allyield}
\end{subfigure}
\hfill
\begin{subfigure}[h]{0.48\linewidth}
        \includegraphics[width=\linewidth]{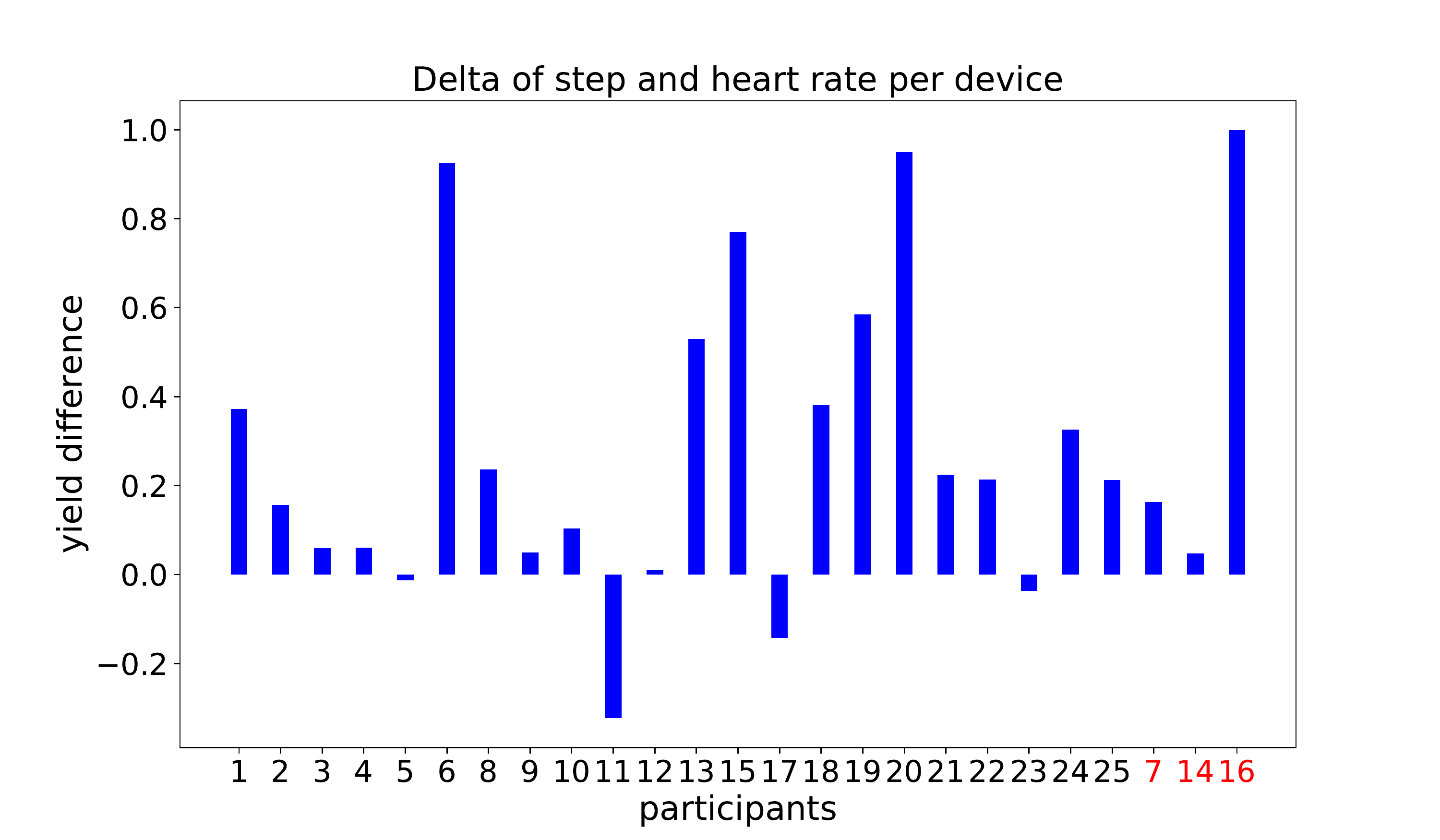}
\caption{}\label{fig:deltahrvstep}
\end{subfigure}
\caption{Yield of heart rate, step and sleep. (a) shows the yield for each participant. The three participants to the right are those with compliant issue. (b) is the difference between step yield and heart rate yield per participant.}
\end{figure}

Figure~\ref{fig:allyield} shows yield of heart rate, step and sleep data for each participant. The average yield of step is much higher than that of heart rate. The mean yield of step count is 0.9113 and the standard deviation is 0.1237, $84\%$ of participants have step yield higher than 0.8, as shown on Figure~\ref{fig:allyield}. In comparison, the mean yield of heart rate is 0.6369 and the standard deviation is 0.3262, $54\%$ participants have heart rate yield higher than $80\%$. Generally, there are three main causes for low yield. The first cause is the user compliance. The patients may not follow the procedure all the time, for instance, they may take off Fitbit occasionally. The second cause is the sensor issue. There may be the case where the sensor is not working. The third cause is the lose of connectivity. Fitbit cloud may permanently lose some data due to bad network connectivity. If the end-to-end latency is longer than the period when Fitbit store the unsynchronized data locally, we will permanently lose these data since they will be overwritten by the later coming data. In section 4.3, we will show that this is not the cause of low yield for our system.

For those users which have either one of the heart rate yield and step count yield higher than $0.8$, the results demonstrate that the users are highly compliant with the study protocol, since we can verify that the Fitbit is continuing collecting the data. As shown on Figure~\ref{fig:allyield}, $88\%$ of participants are compliant with the study protocol. However, for those users, such as 7, 14 and 16, which have both low heart rate yield and step yield, the results imply the low user compliance. There is also the case where the yield of one sensing modality is significantly higher than the other. For example, Figure~\ref{fig:deltahrvstep} show that user 6, 13, 15, 16, 19, 20 have much higher step yield than their heart rate yield. On the other hand, only user 11 has heart rate yield which is significantly larger than its step yield. The inconsistency of heart rate yield and step yield can reflect the productivity and reliability of different sensing modality. From the yield analysis, we can conclude that Fitbit's step measurement is usually more productive and reliable than heart rate measurement. The reason of relatively low heart rate yield may be explained by the sensing mechanism. Step is derived from acceleration which is collected by Inertial Measurement Unit (IMU) inside Fitbit. Fitbit measure heart rate via shooting LED light to skin which detects changes in blood volume. Thus, a period of blood volume changing indicates a single heart beat. We can imagine that IMU can continuously collect acceleration while Fitbit turning on, no matter how Fitbit is worn. However, the heart rate sensor more relies on the proper wearing position. From its sensing mechanism, we can see that heart rate sensor works well only if it is tightly fasten to user's wrist.

The poor result of sleep yield, Figure~\ref{fig:allyield}, indicates that the sleep measurement is not as reliable as step count and heart rate. The Fitbits of user 14, 16, 19, 21 don't have any sleep measurement at all. Fitbit Charge HR can automatically detect user's sleep behavior via a combination of movement and heart-rate patterns. Study ~\cite{Zambotti17} point out that sleep stages can not be returned in three cases, such as the heart rate can not be clearly detected throughout the night, sleep duration is less than three hours or battery runs out of power during the sleeping period. Despite the fact that all sleep yield is low, we still want to distinguish the sleep yield as relatively high and relatively low. In the analysis, we define the relatively high sleep yield as its value is larger or equal to the median, thus the relatively low sleep yield are those less than the median. The Fitbits of user 6, 7, 14, 15, 16, 19, 21, which have no or relatively small sleep yield are those also have low heart rate yield. The reason may be that users are unwilling to wear Fitbit during the sleep. Some of them, such as user 6, 15, 19, 21, they have high step yield over 0.8, which indicates that they are compliant with protocol of wearing Fitbit during the day time. The users 7, 14, 16, who give very low sleep yield are those with both low heart rate and step yield. Therefore, the participants associated with these three devices are less compliant throughout the day. Since the step sensing is always generating data even if the user take off the device, the only cause of low step yield is that the device is out-of-battery. 

The yield analysis from different sensing modalities demonstrates the ability of analyzing user compliance issue. Furthermore, the differences of yield from sensing modality suggest the reliability of associated sensors. In section 4.2, we will further discuss other metrics to measure reliability.

\subsection{Reliability}

\begin{figure} [h!]
\begin{subfigure}[h]{0.48\linewidth}
		\includegraphics[width=\linewidth]{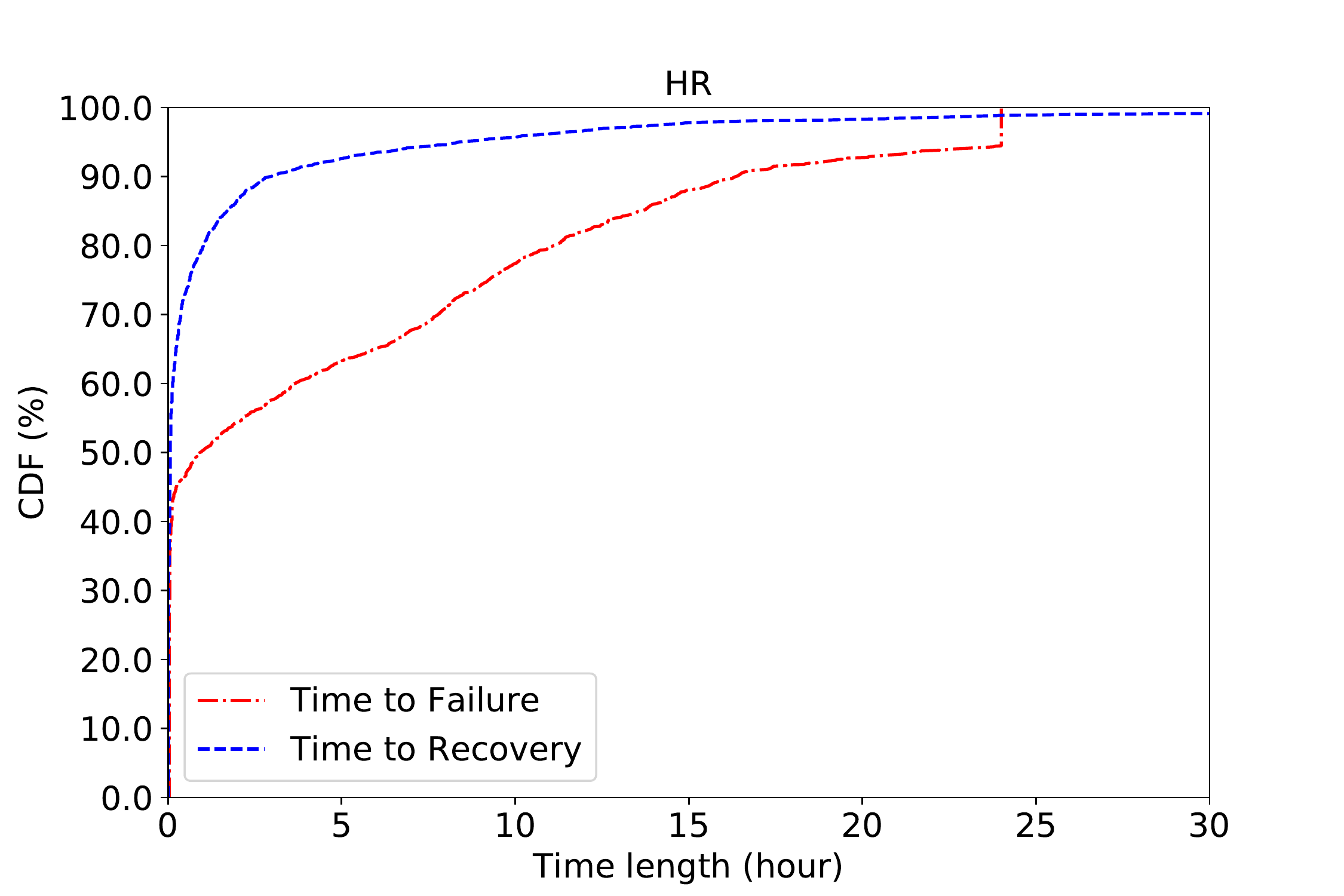}
\caption{}\label{fig:hr_cdf}
\end{subfigure}
\hfill
\begin{subfigure}[h]{0.48\linewidth}
        \includegraphics[width=\linewidth]{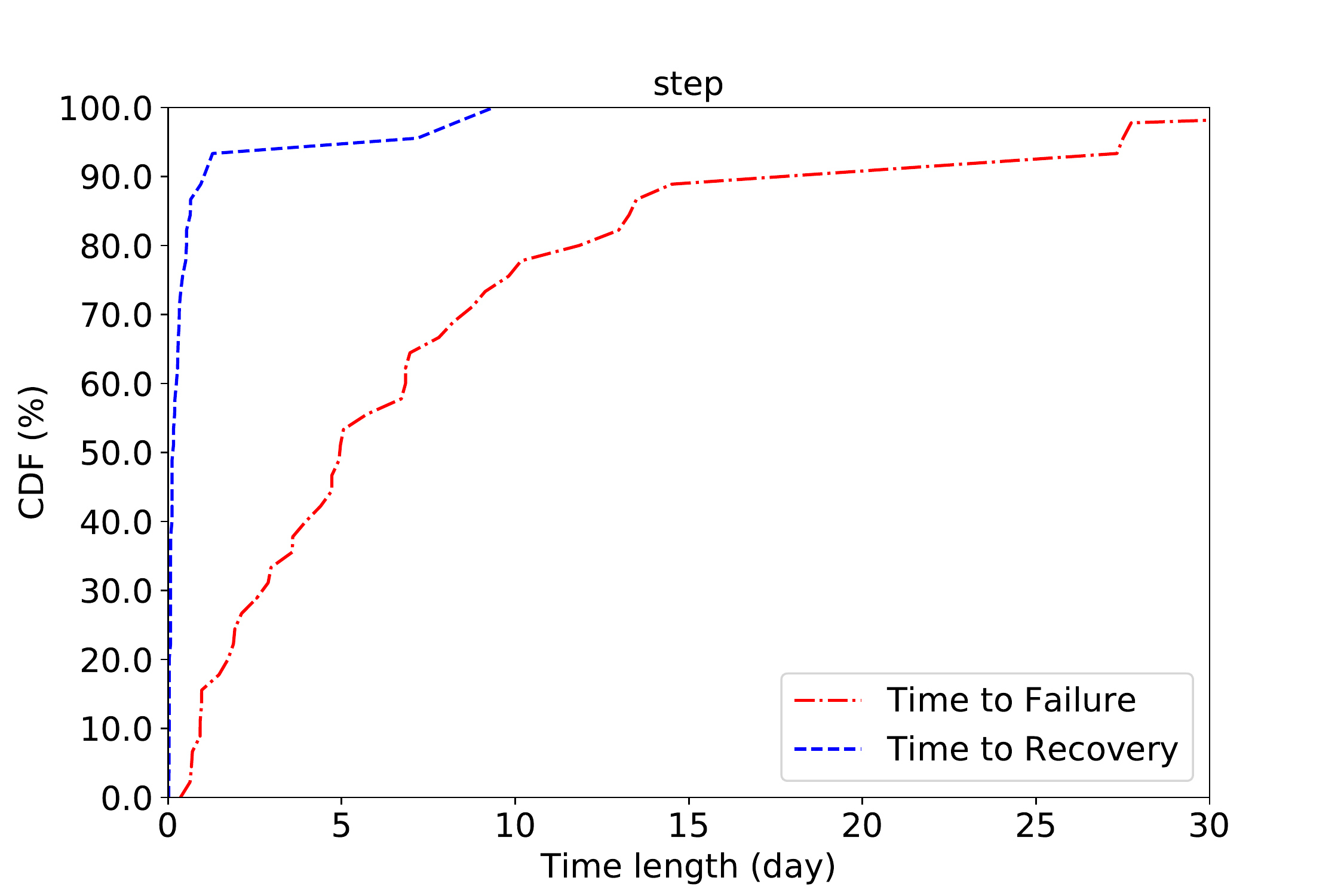}
\caption{}\label{fig:step_cdf}
\end{subfigure}
\caption{CDF of time-to-failure and time-to-recovery. (a) shows the CDF of time-to-failure and time-to-recovery of heart rate sensing. (b) shows the CDF of time-to-failure and time-to-recovery of step sensing.}
\end{figure}

In this section, we will explore two intuitive metrics measuring system's reliability. Time-to-failure is defined as the time interval during which a component operates continuously without a failure and time-to-recovery is defined as the time interval from the occurrence of a failure until the component recovers \cite{Chipara2010}. Time-to-failure measures how frequently our system fail, on the other hand, time-to-recovery measures how quickly our system is able to recover from failure. In our system, failure is the case where a certain type of data is missing. We analyze the system reliability in terms of heart rate and step data instead of sleep data, since heart rate and step are continuously collected by Fitbit at every minute, which can better reflect system reliability through the entire duration of study. Time-to-failure is calculated as the length of duration when there is no data missing. Time-to-recovery is actually the length of gap where the data is missing. Step has an average of $0.035$ times of failure per participant per day, which is significantly less than that of heart rate, $3.126$ times of failure per participant per day. Thus, step sensing is less likely to fail than heart rate sensing. The median time-to-failure of heart rate sensing is 55 minutes, and the median time-to-failure of step sensing is 119 hours (about 5 days). However, when looking at time-to-recovery, heart rate sensing recovers from failure more quickly than step sensing. The median time-to-recovery of heart rate sensing is 4 minutes, and the median time-to-recovery of step sensing is 198 minutes (3.3 hours). Heart rate sensing usually takes less time to recovery, which can be observed when comparing Figure~\ref{fig:hr_cdf} and Figure~\ref{fig:step_cdf}. Usually, we care about the tail of time-to-recovery, since it can tell us about whether our system can handle the extreme case of long duration of failure. Figure ~\ref{fig:hr_cdf} and Figure~\ref{fig:step_cdf} plot the CDF of time-to-recovery for all devices. The $95\%$ percentile time-to-recovery of heart rate is 8.4 hours and 137 hours (5.7 days) for step. Although the extreme case of long time-to-recovery of step sensing happen very seldom, it is still disconcerting for the real clinical situation, since the clinicians may miss the appropriate time to provide just-in-time intervention if the system failure last for a long period.

\subsection{Latency}

\begin{figure}
		\center
		\includegraphics[scale=0.38]{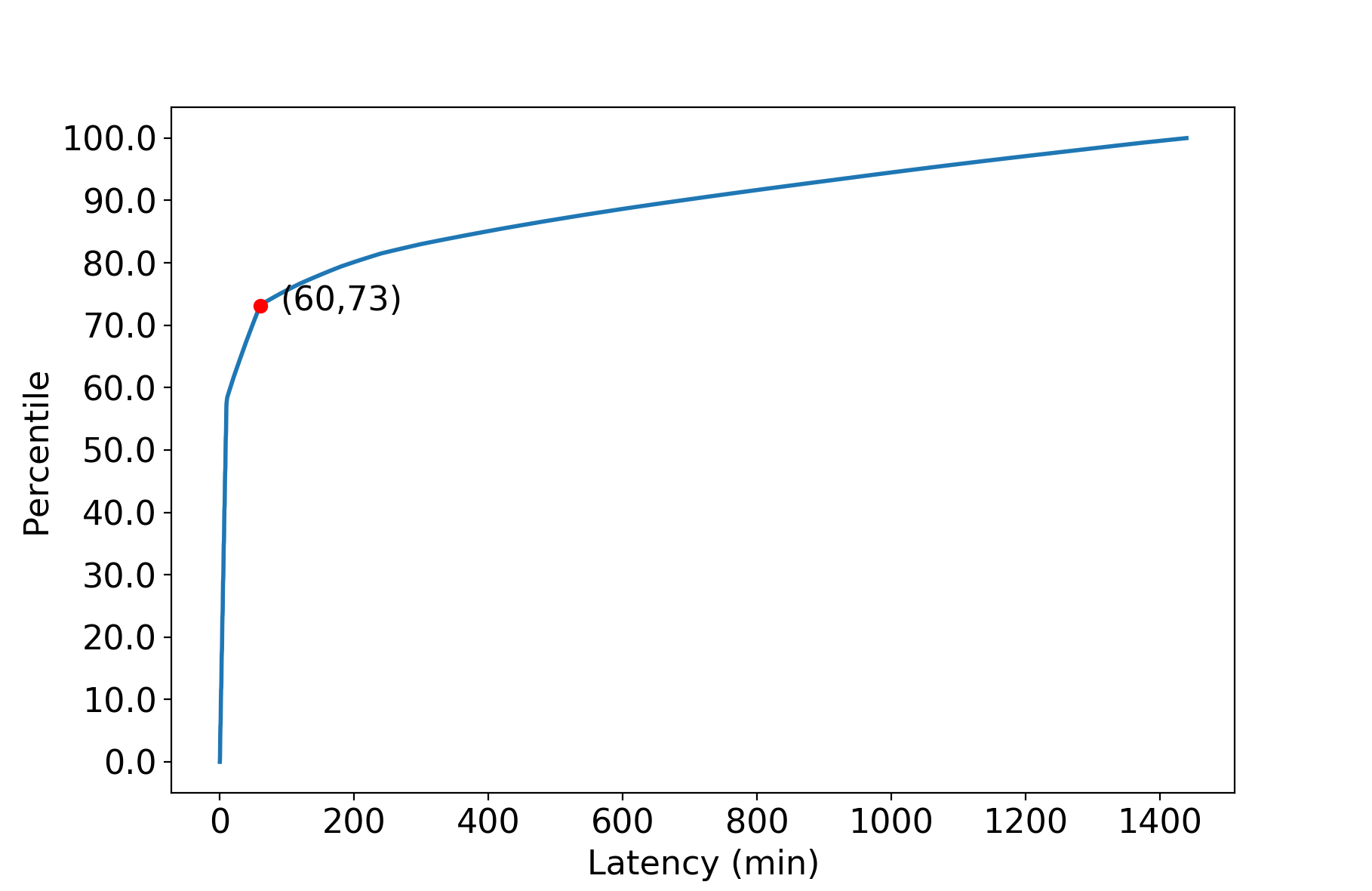}
        \caption{CDF of system latency}\label{fig:latency_one}
\end{figure}

A well-functioning system need to have high yield and high reliability as well as low end-to-end latency. Studying the device's latency is important as it provides us with a good idea of whether or not the system can be used in a time-sensitive manner. In our analysis, the end-to-end system latency is defined as the time duration from when the data is collected by Fitbit to the time of it is synchronized with Fitbit cloud. Since Fitbit synchronizes all types of data simultaneously, we can analyze end-to-end system latency in terms of one type of data. Therefore, step data is selected to measure latency because of its highest reliability. We plot the cumulative distribution function (CDF) of the end-to-end latency, which can be seen in Figure~\ref{fig:latency_one}. From the latency result analysis, we learn the fact that the median end-to-end system latency is 8.55 minutes and $99\%$ percentile of latency is 22.5 hours, which is less than a day. As Fitbit can locally store data up to 7 days, the end-to-end system is not an issue of causing data loss. Also, the CDF tells us $73\%$ of data transmission is under an hour, meaning that our system is capable of quickly responding to patient's deterioration in most of time. We would like to mention that such low latency is not necessary in this particular study. But the latency analysis explore the potential for other applications that require timely alerts and intervention. Our data collection system is feasible of applying to hourly real-time monitoring and just-in-time intervention. The latency can be further reduced if the system is implemented natively in Fitbit Cloud.

\section{Data Preprocessing and Feature Selection}
\subsection{Data Preprocessing}
Since Fitbit API has its own data preprocessing protocols, we could only obtain the processed data provided by Fitbit cloud. Time series data from Fitbit is preprocessed as summary given for a certain granularity. In the study, time granularity for heart rate, step count and sleep status is chosen as one minute. Sleep summary data is generated by Fitbit API for each sleep duration. Since patients are supposed to wear Fitbit all the time during the study period, there is inevitably some noisy data unrelated to the features we want to extract. Step count is continuously measured by Fitbit through the whole day, even when patients go to bed. Since our purpose of measuring step count is to reflect patient's daily activity level when they awake, a time filter is applied to extract step count for the time period when patient is "awake". Sleep time is defined as the time when total step count is under 10 within 30 minutes after 7pm. Awake time is defined as the time when the first step is taken after 7am \cite{Bae2016}. Thus, we only consider steps made within the duration between Awake time and Sleep time. 

Particularly, there are four devices missing sleep data. In order to preserve the consistency of features among training examples, we fill in the average value of corresponding missing features calculated from other examples which have complete features. 
\subsection{Features}
\subsubsection{First Order Features}
First order statistical features used in our project are mean, maximum, minimum, skewness and kurtosis. Assume given time series $\{x_i\},1\leq i \leq N$, mean $\mu$ and standard deviation $\sigma$ of time series data is calculated as:
\begin{equation}
\begin{aligned}
\mu = \frac{\sum_{i=1}^Nx_i}{N}, \sigma = \sqrt{\frac{\sum_{i=1}^N(x_i-\mu)}{N}}
\end{aligned}
\end{equation}
Skewness is defined as:
\begin{equation}
\begin{aligned}
Skewness = \frac{\sum_{i=1}^N(x_i-\mu)^3}{(N-1)\sigma^3}
\end{aligned}
\end{equation}
which measures the symmetry of distribution. If skewness is high, it means the distribution lacks symmetry. Skewness can help us model whether the time series data, such as heart rate and sleep data, follows a bell-shape distribution. Kurtosis, in the other hand, measures whether the distribution is heavy-tailed or light-tailed compared to normal distribution. Kurtosis can be calculated using following formula:
\begin{equation}
\begin{aligned}
Kurtosis = \frac{\sum_{i=1}^N(x_i-\mu)^4}{(N-1)\sigma^4} - 3
\end{aligned}
\end{equation}
\subsubsection{Second Order Features}
The commonly used second order time series features in medical data mining are co-occurance features \cite{Wang17, Mao2012}. Those features are shown to outperform other second order features in the case of one dimensional time series data \cite{Mao2012}. First of all, one dimensional time series data is quantized into Q levels. Then a two dimensional matrix $c(i,j), (1\leq i,j\leq Q)$ is created by calculating number of times two points of quantized level $i$ and $j$ is at distance $d$. Second order features, such as energy ($E$), entropy ($S$), correlation ($\rho_{x,y}$), inertia ($F$) and local homogeneity (LH) are derived based on matrix $c$: 
\begin{equation}
\begin{aligned}
E &= \sum\limits_{i=1}^Q\sum\limits_{j=1}^Qc(i,j)^2 \\
S &= \sum\limits_{i=1}^Q\sum\limits_{j=1}^Qc(i,j)\ast \log(c(i,j)) \\
\rho_{x,y} &= \frac{\sum_{i=1}^Q\sum_{j=1}^Q(i-\mu_x)(j-\mu_y)c(i,j)}{\sigma_x\ast\sigma_y} \\
\end{aligned}
\end{equation}
where:
\begin{equation}
\begin{aligned}
\mu_x = \frac{\sum_{i=1}^Qi\sum_{j=1}^Qc(i,j)}{Q},~~~~ \sigma_x^2 = \frac{\sum_{i=1}^Q(i-\mu_x)^2\sum_{j=1}^Qc(i,j)}{Q} \\
\mu_y = \frac{\sum_{j=1}^Qj\sum_{i=1}^Qc(i,j)}{Q},~~~~ \sigma_y^2 = \frac{\sum_{j=1}^Q(j-\mu_y)^2\sum_{i=1}^Qc(i,j)}{Q} \\
\end{aligned}
\end{equation}
\begin{equation}
\begin{aligned}
F &= \sum\limits_{i=1}^Q\sum\limits_{j=1}^Q(i-j)^2c(i,j) \\
LH &= \sum\limits_{i=1}^Q\sum\limits_{j=1}^Q\frac{1}{1+(i-j)^2}c(i,j)
\end{aligned}
\end{equation}
\subsubsection{Detrended Fluctuation Analysis}
In stochastic processes, chaos theory and time series analysis, detrended fluctuation analysis (DFA) is a method for determining the statistical self-affinity of a signal \cite{PhysRevE.49.1685,PhysRevE.64.011114}. Self-affinity is an important factor when analyzing time series which is supposed to have a regular patten. In our case, we apply DFA to heart rate and sleep time series, which evaluates  long-range correlation of noisy time series data \cite{PhysRevE.64.011114}. DFA can be used for analyzing non-stationary time series with slowly varying trend, such as heart rate \cite{Mao2012}. 

DFA is the average fitting error over time series segments of different scale \cite{Mao2012}. It first convert time series $\{x_i\},1\leq i \leq N$ into an unbounded process $\{X_j\},1\leq j \leq N$ by summation or integration:
\begin{equation}
\begin{aligned}
X(j) = \sum\limits_{i=1}^j[x(i)-\langle x \rangle], 1 \leq j \leq N
\end{aligned}
\end{equation}
where $\langle x \rangle$ is the mean over the entire time series $\{x_i\},1\leq i \leq N$. Then $\{X_j\},1\leq j \leq N$ is divided into sub-series each of length $n$. A polynomial piecewise fit $\{Y_j\},1\leq j \leq N$ is generated by minimizing local least square error within the sub-series. The fluctuation ($F$) of detrended time series is:
\begin{equation}
\begin{aligned}
F(n) = \sqrt{\frac{1}{N}\sum\limits_{j=1}^N(X(j)-Y(j))^2}
\end{aligned}
\end{equation}
Finally, fluctuation measure is repeated over different length $n$.
\subsubsection{Feature Extraction}
\begin{figure}
		\center
		\includegraphics[scale=0.4]{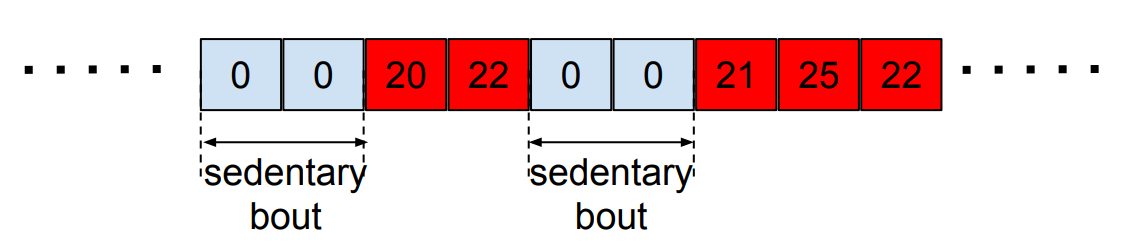}
        \caption{Sedentary bout is the duration when there is 0 step count. Each block represent 1 minute granularity. }\label{fig:step_illu}
\end{figure}

\begin{figure}
		\center
		\includegraphics[scale=0.4]{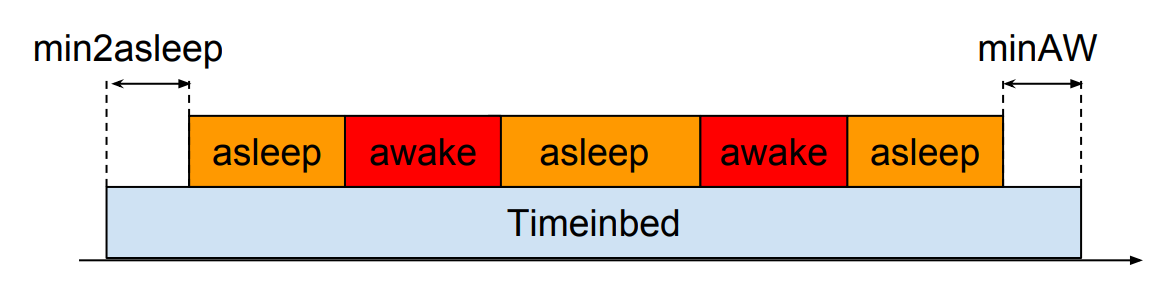}
        \caption{Illustration of sleep parameters. Basically, TimeinBed is the sum of minutesToFallAsleep, minutesAsleep, minutesAwake, and minutesAfterWakeup.}\label{fig:slepp_illu}
\end{figure}

Fitbit API preprocessed raw data collected from Fitbit Charge HR. Various calculations are performed to extract meaningful features from time series data. Instead of using time series data directly obtained from Fitbit API, we transform them into statistical features. Also, we derive other features from the original step data. Sedentary bout time is defined as a time duration where no steps are taken when patient is awake, as illustrated by Figure~\ref{fig:step_illu}. Daily sedentary bout count refers to the total number of sedentary bout per day. Statistical features, such as min, max, and mean, are calculated from original step data and the derived sedentary behavior data. Sleep data gathered by Fitbit includes sleep status and sleep summary. Sleep status takes discrete value from 0 to 3, where 0 is no measurement, 1 is sleeping, 2 is restless, 3 is awake. Sleep status is measured every minute as a time series data. First order time series features, skewness and kurtosis, are extracted as input to train models. We apply DFA to sleep status, an indication of sleep quality where high value means more fluctuation in sleep. Different time windows are generated to compensate the influence of unknown appropriate window size. Sleep summary data provides various features which are extracted automatically by Fitbit API. The useful features are Time in Bed, Minute to Fall Asleep, Minute Awake, Minute After Wakeup, Awake Count, Restless Count, Restless Duration. The relation of the parameters is illustrated by Figure~\ref{fig:slepp_illu}. We then compute min, max, mean of those features. Since we work on predicting deterioration for heart failure patients, time series analysis is also performed on heart rate data. First order time series features, such as mean, standard deviation, skewness and kurtosis, are derived from heart rate data of each patient. Second order time series features, such as energy, correlation, inertia, local homogeneity, are also calculated. We perform DFA on heart rate data, which will capture the heartbeat fluctuation along the day. Different time windows are also applied for extracting DFA of heart rate. There are 51 features in total, representing characteristics of activity, sleep and heart rate. In order to improve model performance, we apply Sequential Forward Feature Selection to filter out irrelevant features. Thus, a subset of overall features will be used to train a predictive model.

\section{Deterioration Prediction}
Due to the study scale and the challenge of launching study for heart failure patients, we finally get 25 patients participating the study and 7 of them deteriorated within 60 days. Typically, this amount of examples is insufficient to train a stable predictive model. Moreover, we are facing the issue of incomplete data and noisy measurement. Hence, we need to modify the standard machine learning algorithms to robustly learn a predictive model and overcome the overfitting issue brought by small data set. In the following section, we will focus on solving two predictive tasks which are commonly studied in the clinician research. One is to train a model predicting the deterioration happening in the future with the data collected by sliding window, namely the early warning task. The other is to predict whether a patient will deteriorate in the future based on the data collected from the beginning of monitoring, namely identifying the patients with the risk of deterioration. Since the deterioration only happen in a small fraction of days during the whole monitoring period, the challenge in the early warning task is to learn a classifier from skewed data set. For identifying the patients with risk of deterioration, the challenge is the size of dataset, because each patient's data is an example.
\subsection{Deterioration Early Warning}
\subsubsection{Dataset}
In the deterioration early warning task, the dataset is created by treating each day as a training example. The feature extraction methods are performed to segment the time series data and generate meaningful statistics features in one day granularity. This is a binary classification task with normal day and deterioration day as the classes. After segmenting the time series data, we obtain a dataset of 427 normal days and 11 deteriorated days. The classification task is highly skewed and one of the class has very small amount of samples. Thus, we formalize the learning task as anomaly detection.
\subsubsection{Weighted Samples One Class SVM}
One Class SVM (OC-SVM) is a semi-supervised machine learning method, which is designed for anomaly detection. OC-SVM is proposed assuming the anomaly examples are not available during the training phase. Therefore, it is trained with only normal examples. However, SVM classifiers are sensitive to outliers in the training examples~\cite{Xu2006} and need outlier suppression techniques in the training phase. Hence, we introduce the sample weights which are learned during the training phase to assign each training example a weight $\eta_i$ to control its influence on the decision boundary. Therefore, the revised optimization objective of weighted samples OC-SVM is: 
\begin{equation}\label{eq:org}
\begin{gathered}
\min\limits_{\mathbf{\omega}, \rho, \mathbf{\eta}} \frac{\left \| \omega \right \|^2}{2} - \rho + \frac{1}{\nu n} \sum\limits_{i=1}^n \eta_i \max(0,\rho-\mathbf{\omega}^T\phi(x_i)) \\
\text{s.t. ~}e^T \eta \geq \beta n
\end{gathered}
\end{equation}
where $\nu$ is the hyperparameter to control the upper bound of the fraction of training errors and the lower bound of the fraction of support vectors, $\beta$ controls the maximum number of points allowed to be outliers. The above optimization is not jointly convex in terms of $\omega$, $\rho$ and $\eta$. The non-convex part $\frac{1}{\nu n} \sum\limits_{i=1}^n \eta_i \max(0,\rho-\mathbf{\omega}^T\phi(x_i))$ can be reformulated using concave duality~\cite{Zhang08,Amer2013}. Then the above objective~(\ref{eq:org}) is equivalent to:
\begin{equation}
\begin{gathered}
\min\limits_{\omega,\rho} R_{vex} + R_{cave} \\
R_{vex} = \frac{\left \| \omega \right \|^2}{2} - \rho \text{,~} R_{cave} = \min\limits_{\eta} \sum\limits_{i=1}^n \eta_i \max(0,\rho-\mathbf{\omega}^T\phi(x_i)).
\end{gathered}
\end{equation}
Obviously, $R_{vex}$ is a convex objective, and $R_{cave}$ is a non-convex objective. We reformulated $R_{cave}$ as:
\begin{equation}
\begin{gathered}
R_{cave} = g(h(\omega)) \\
\text{where~~} h(\omega) = \max(0,\rho-\omega^T\phi(x))\text{,~} g(u)=\inf_{\eta \in \{0,1\}} [\eta^T u]
\end{gathered}
\end{equation}
Since $h(\omega)$ is the point-wise infimum of a set of linear functions, $g(h(\omega))$ is concave on the domain $h(\omega) \in \Omega$~\cite{Zhou2011}. We can approximate $R_{cave}$ by the multi-stage relaxation proposed by~\cite{Zhang08}. The basic idea of multi-stage relaxation is to first set $\eta$ to be a vector of ones, indicating all the training examples have the same weights. Then the procedure iteratively minimize the objective by alternatively fixing $\omega, \rho$ and $\eta$ until convergence: \\
\begin{enumerate}
\item Fix $\eta=\hat{\eta}$ and calculate $\hat{\omega}$ via solving
\begin{equation}
\begin{gathered}
\hat{\omega} = \text{argmin}_{\omega} \frac{\left \| \omega \right \|^2}{2} - \rho + \frac{1}{\nu n} \sum\limits_{i=1}^n \hat{\eta_i }\max(0,\rho-\mathbf{\omega}^T\phi(x_i))
\end{gathered}
\end{equation}
\item Fix $\omega = \hat{\omega}$ and calculate $\hat{h}(\hat{\omega}) = \max(0,\rho-\hat{\omega}^T\phi(x_i))$. In order to minimize the overall objective, $\eta$ will be $1$ for $\beta n$ number of smallest $h_i$ in $\hat{h}(\hat{\omega}) $ based on the constrain $e^T \eta \geq \beta n$. \\
\end{enumerate}
The weighted samples OC-SVM begins with the standard OC-SVM by initializing the sample weights to be all ones. The multi-stage optimization framework could obtain a better solution than the standard OC-SVM after convergence~\cite{Zhou2011}. In the experiment section, we will compare the weighted samples OC-SVM with the standard one as well as other anomaly detection models to demonstrate its superior performance for solving the deterioration early warning problem.
\subsection{Deterioration Risk Prediction}
\subsubsection{Dataset}
In the deterioration risk prediction task, the dataset consists of the derived features from deteriorated participants and non-deteriorated participants, which are extracted from the beginning of the monitoring. There are totally 25 examples, including 7 examples as the patients who deteriorated within 60 days and 18 examples as patients who did not deteriorate within 60 days. Note that the dataset here is extremely small and imbalanced. Most of the models will overfit when applied to solve machine learning problem given such small dataset. In order to achieve high accuracy as well as preventing overfitting, we use $K$ Nearest Neighbor to perform the binary classification.

\subsubsection{K Nearest Neighbor}
$K$ Nearest Neighbor (KNN) is a non-parametric learning method for classification. An example is classified by the majority vote of its $K$ nearest neighbors. Usually, $K$ is predefined before training. In the experiment, we will explore the best $K$ for this specific task. The distance metric used in our implementation is the Euclidean distance of the two sample data which takes into account all available features. Also, we will perform feature selection on the training set to find the best feature subset which could achieve better testing accuracy.

\section{Performance Evaluations}
We evaluate the performance of proposed models on the real experimental data collected from heart failure patients in a large research hospital in United States. The clinical study is IRB approved and there are a total of 25 patients recruited for research purposes on predicting clinical deterioration. In the study, patients were supposed to wear Fitbit Charge HR to collect health related data everyday, such as heart rate, step and sleep data. The data was finally transmitted to our Heroku database. 5 out of 25 patients became deteriorated (readmitted or deceased) within 30 days after being discharged from hospital. 2 patients became deteriorated during the period between 30 days and 60 days. The patients are mixed in age ranging from 66 to 88, 16 male vs. 9 female, and LACE index ranging from 3 to 15.

\subsection{Deterioration Early Warning}
There are a total of 438 days which have valid data collected from the Fitbit devices, including 427 normal days and 11 days when deterioration happened. We divide the samples into a training set which contains only normal days and a testing set which contains both normal days and deterioration days. The testing set is designed in such a way to avoid favoring classifier which achieves high accuracy by only predicting the anomaly case for new unseen data. We randomly split the normal days as 95\% of them belonging to the training set and remaining 5\% of them belonging to the testing set. The deterioration days are all counted into the testing set, since OC-SVM only use normal data for training. For each time of evaluation, we repeat the training and testing split for 100 times and average the results. 

\begin{figure}
		\center
		\includegraphics[scale=0.4]{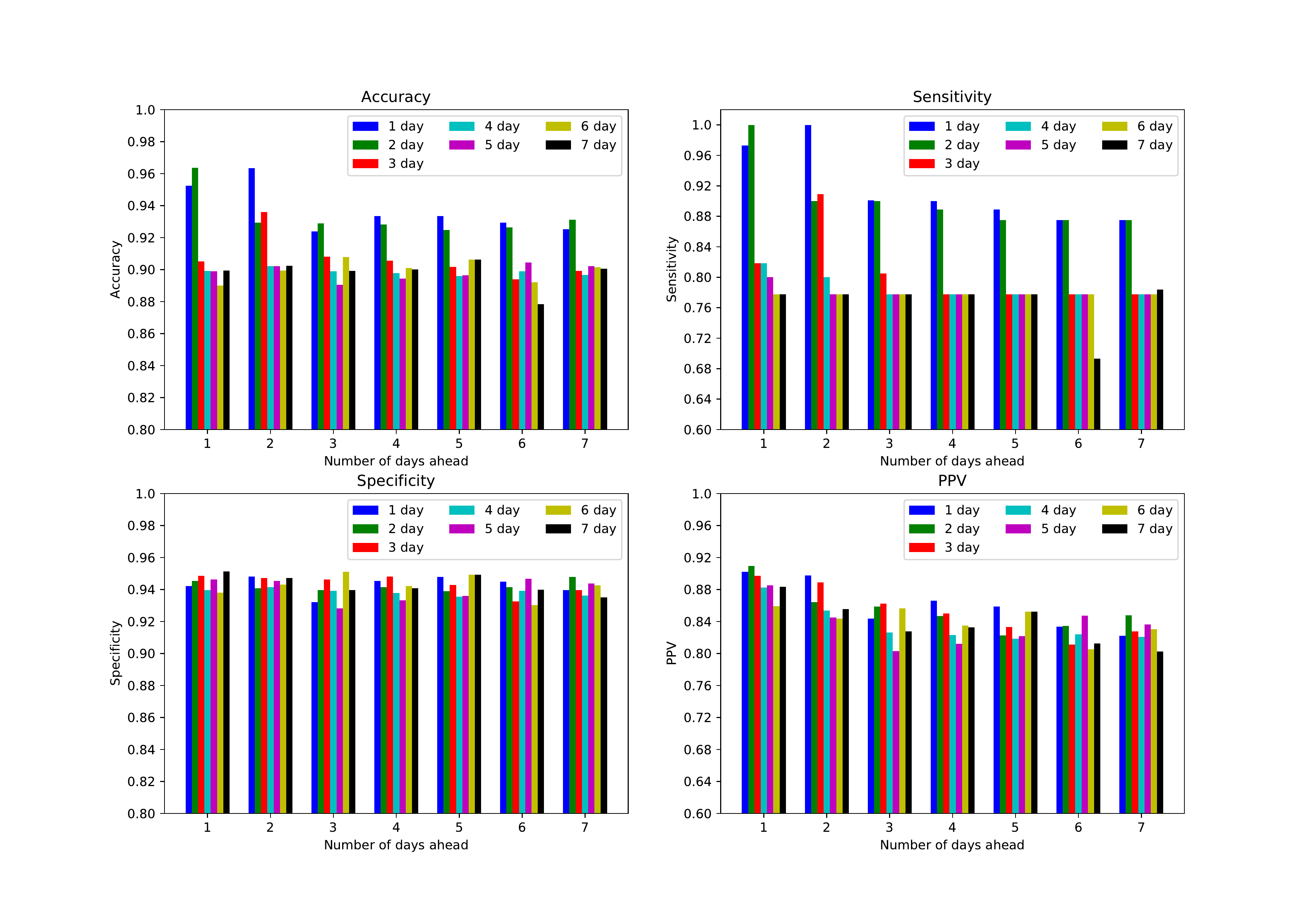}
        \caption{Performance evaluation of weighted samples OC-SVM with varying size of time window and number of days ahead. The classifier is able to achieve good performance for predicting near future deterioration happening.}\label{fig:svm_plot}
\end{figure}

We first explore the predictive performance of different combination of time window size and the days ahead. Since there is a lack of empirical guidelines for how to choose the time window and how well the predictor performs for more days ahead. It is important to find the relationship between time window size and the number of days ahead for prediction. In the experiment, we train weighted samples OC-SVM with time window varying from 1 to 7 days for predicting deterioration ahead by 1 to 7 days. The results are shown on Figure~\ref{fig:svm_plot} with x-axis indicating the number of days ahead. For each number of days ahead, we compare the performance metrics by varying time window size, which are indicated by bars of different colors. The accuracy and sensitivity are high for predicting the deterioration when just using time window of 1 or 2 days. The combination of predicting 1 day ahead using 2 days data achieves the highest accuracy of $0.9636$ and also highest sensitivity of $1.0$ with specificity of $0.9480$ and PPV of $0.8975$. The results imply the performance declines when predicting more days ahead, which is reasonable since the long-term future may not be related to current status. To our surprise, the weighted samples OC-SVM perform better when the window size is small. The model is able to achieve over $0.9$ overall accuracy when just using 1-3 day data. The results give us a practical guideline of using less than 3 day data to predict the deterioration in the near future.
 
\begin{figure}
		\center
		\includegraphics[scale=0.4]{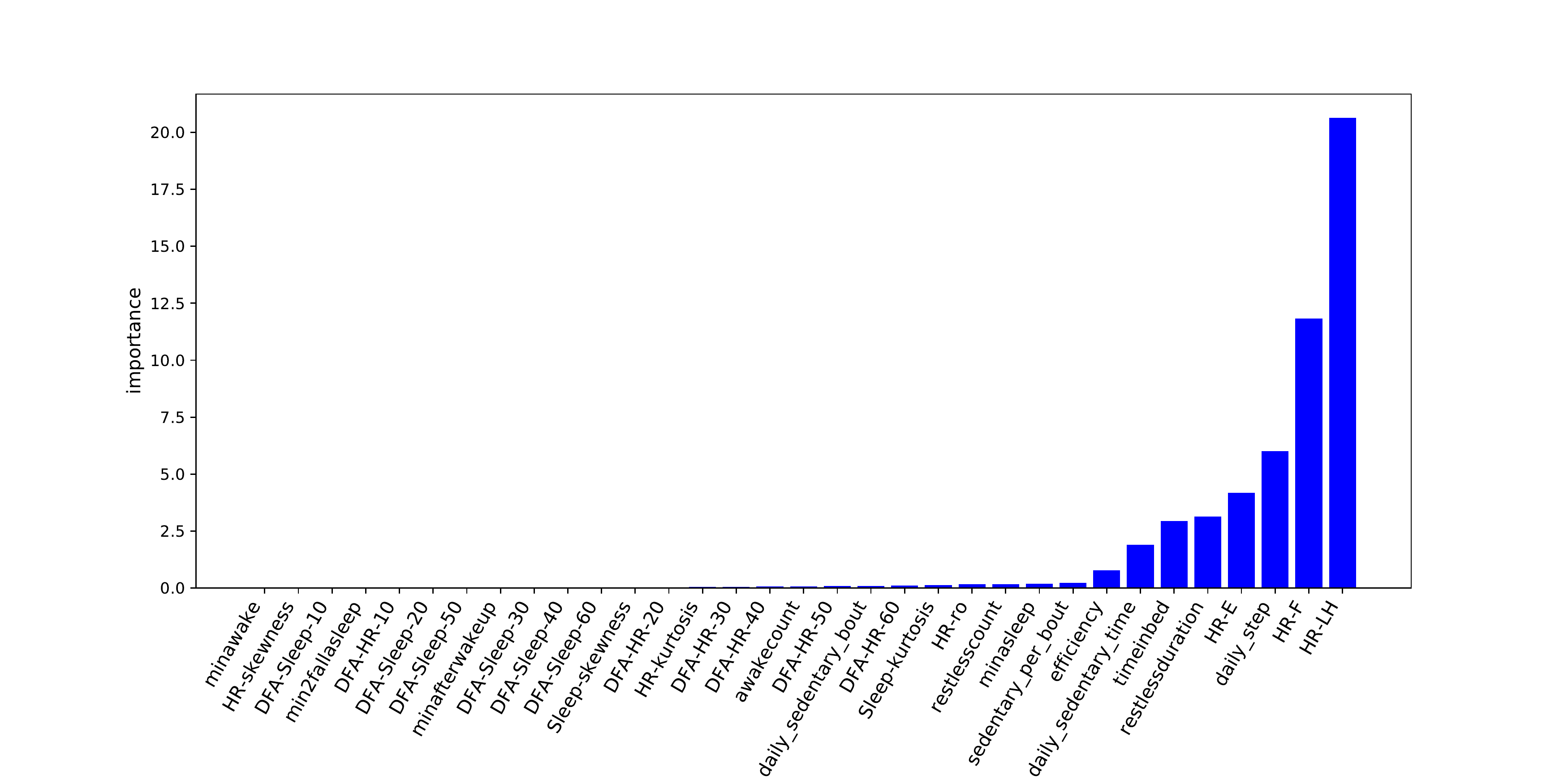}
        \caption{Feature importance of weighted samples OC-SVM. Heart rate features contribute significantly to the classification results. Sleep and step features also have impact on the classification.}\label{fig:fi}
\end{figure}

Then, we compare weighted samples OC-SVM with other commonly used methods for anomaly detection, including standard OC-SVM, density based method such as local outlier factor (LOF) and clustering based method such as $K$ means clustering. The dataset here is the same as that used in the last experiment. In this particular experiment, we set the time window to be 2 days and predict whether the participant will deteriorate after 1 day. We evaluate the model performance via looking at accuracy, sensitivity, specificity and precision. Table~\ref{tab:ad_results} summarizes the results for different anomaly detection methods. The $K$ means and LOF have low sensitivity compared with OC-SVM based approach, which imply these two methods are not suitable for solving our specific problem. Our proposed weighted samples OC-SVM performs better than the standard OC-SVM with lower false positive rate. Figure~\ref{fig:nu_roc} demonstrates the influence of $\nu$ on model's performance. $\nu$ is a hyperparameter of the soft-margin SVM, which controls the upper bound of training error and lower bound of support vectors. The curve serves as the guideline for how to tune the classifier to meet specific performance requirement. For instance, in order to avoid missing any future deterioration, we choose $\nu=0.09$ to achieve the highest sensitivity as well as high specificity. Aside from the performance, we also care about the features that have high influence on the classification results. Figure~\ref{fig:fi} shows the feature importance for each feature used in weighted samples OC-SVM, which is the absolute value of SVM's coefficient. Time series features derived from heart rate, such as local homogeneity, inertia and energy, contribute significantly to the classification results. Besides, step features and sleep features also have impact on the classification results.

These results demonstrate the potential of \textit{early} warning of clinical deterioration to allow just-in-time intervention.

\begin{table}[h!]
\begin{tabular} { | c | c | c | c | c |}
\hline
 Model & Sensitivity & Specificity & PPV & Accuracy\\ \hline
 $K-$means & 0.6555 & 0.7185 & 0.4473 & 0.6997 \\  \hline
 LOF & 0.0909 & 0.9519 & 0.5820 & 0.6959 \\ \hline
 OC-SVM & 1.0 & 0.9015 & 0.8227 & 0.9308 \\ \hline
 Weighted OC-SVM & \textbf{1.0} & \textbf{0.9481} & \textbf{0.8975} & \textbf{0.9635} \\ \hline
\end{tabular}
\caption {Performance evaluation of different anomaly detection methods. The results are averaged over 100 repeated test on randomly split dataset. Weighted OC-SVM has the best performance.} \label{tab:ad_results} 
\end{table}

\subsection{Deterioration Risk Prediction}
\subsubsection{Model Evaluation}
\begin{figure}[htb!]
		\center
		\includegraphics[scale=0.4]{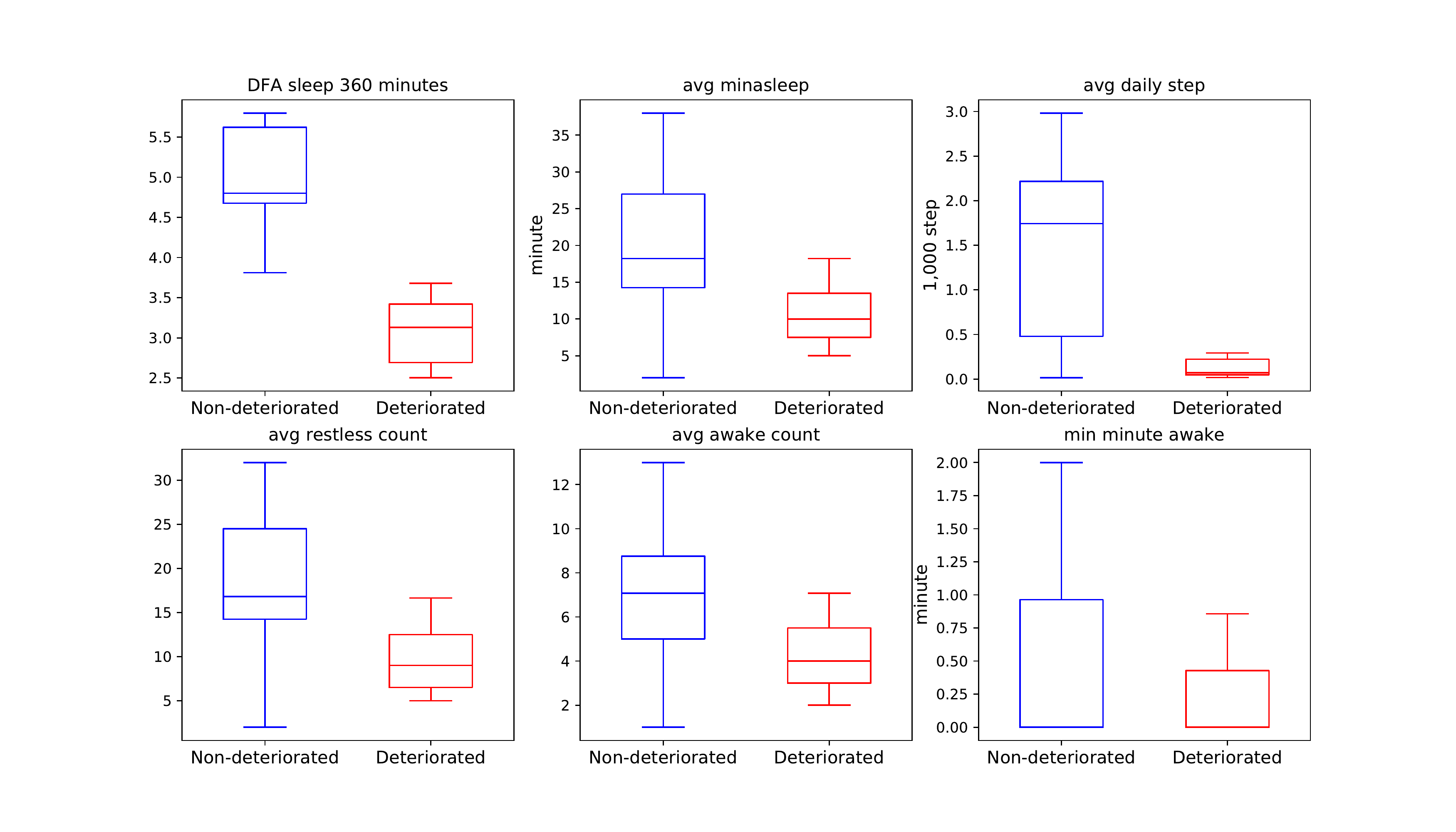}
        \caption{Feature value distribution between non-deteriorated and deteriorated patient groups}\label{fig:box_plot}
\end{figure}

\begin{figure}[htb!]
		\center
		\includegraphics[scale=0.4]{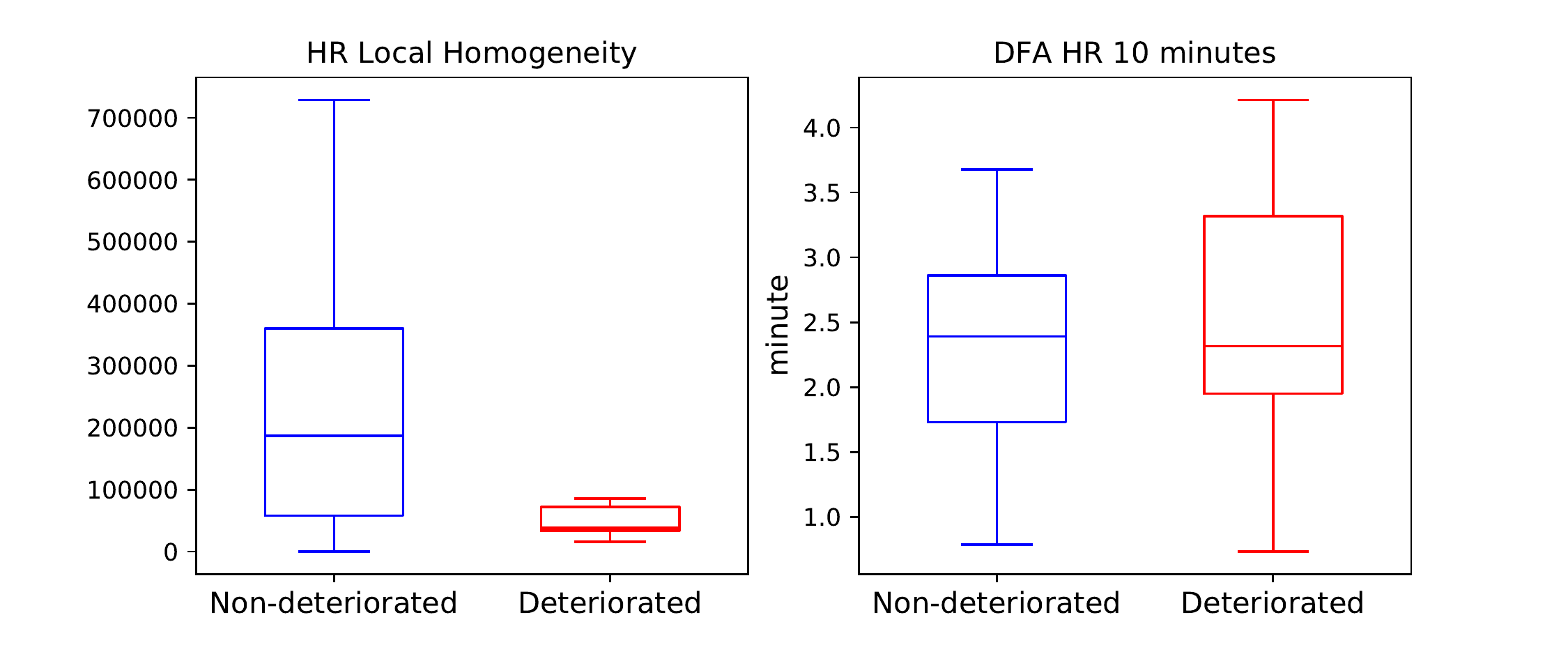}
        \caption{HR feature distribution between non-deteriorated and deteriorated patient groups}\label{fig:box_plot_hr}
\end{figure}

In this section, we aim to predict clinical deterioration risk for each participant within 60 days. Five models including logistic regression, SVM, random forest, neural network and $K$ nearest neighbor, are trained on the data from 25 patients. The performance is evaluated by repeated 5-fold cross validation to tackle the model instability issue induced by the small dataset~\cite{Beleites2008}. We repeatedly perform cross validation for 100 times and the results are averaged among 100 repeated cross validation. The mean results will help eliminate the randomness caused by cross validation. In each iteration of performing cross validation, we apply feature selection on the training set to remove the noisy features and improve prediction accuracy for the testing set. The type of feature selection used in our experiment is the sequential forward feature selection with the accuracy criteria, which selects the best subset of features. The sequential forward feature selection is performed on each model starting with the empty set of features. At each round, a new feature which can improve the overall accuracy is added to the current feature set. Finally, the optimal subset of features is returned by the feature selection algorithm. Table \ref{tab:ft1} shows selected feature subset for different models. We note that all models selected features derived from multiple modalities (step, HR and/or sleep), which suggests the potential benefits of exploiting multi-modal data for predicting clinical deterioration. In particular, random forest, logistic regression and KNN each selected features from all three modalities. There are 6 features which are used by more than two of the models. The features are DFA of sleep using 360-minute window, average minutes of being asleep, average daily steps, average awake counts and minimum minutes of being awake per sleep. Intuitively, we expect the features selected to train predictive models have different distribution among non-readmitted group and readmitted group. Therefore, we plot Figure~\ref{fig:box_plot}, which are the distribution of the 6 most commonly used features. As we expected, the distribution of selected features are different for non-deteriorated group and deteriorated group. Also, we plot the distribution of HR features that were selected by random forest, logistic regression and KNN, shown on Figure~\ref{fig:box_plot_hr}. The distribution of HR local homogeneity is different between non-deteriorated and deteriorated groups. However, the distribution of DFA HR with 10-minute window is similar among the two groups.

\begin{figure}[htb!]
		\center
		\includegraphics[scale=0.42]{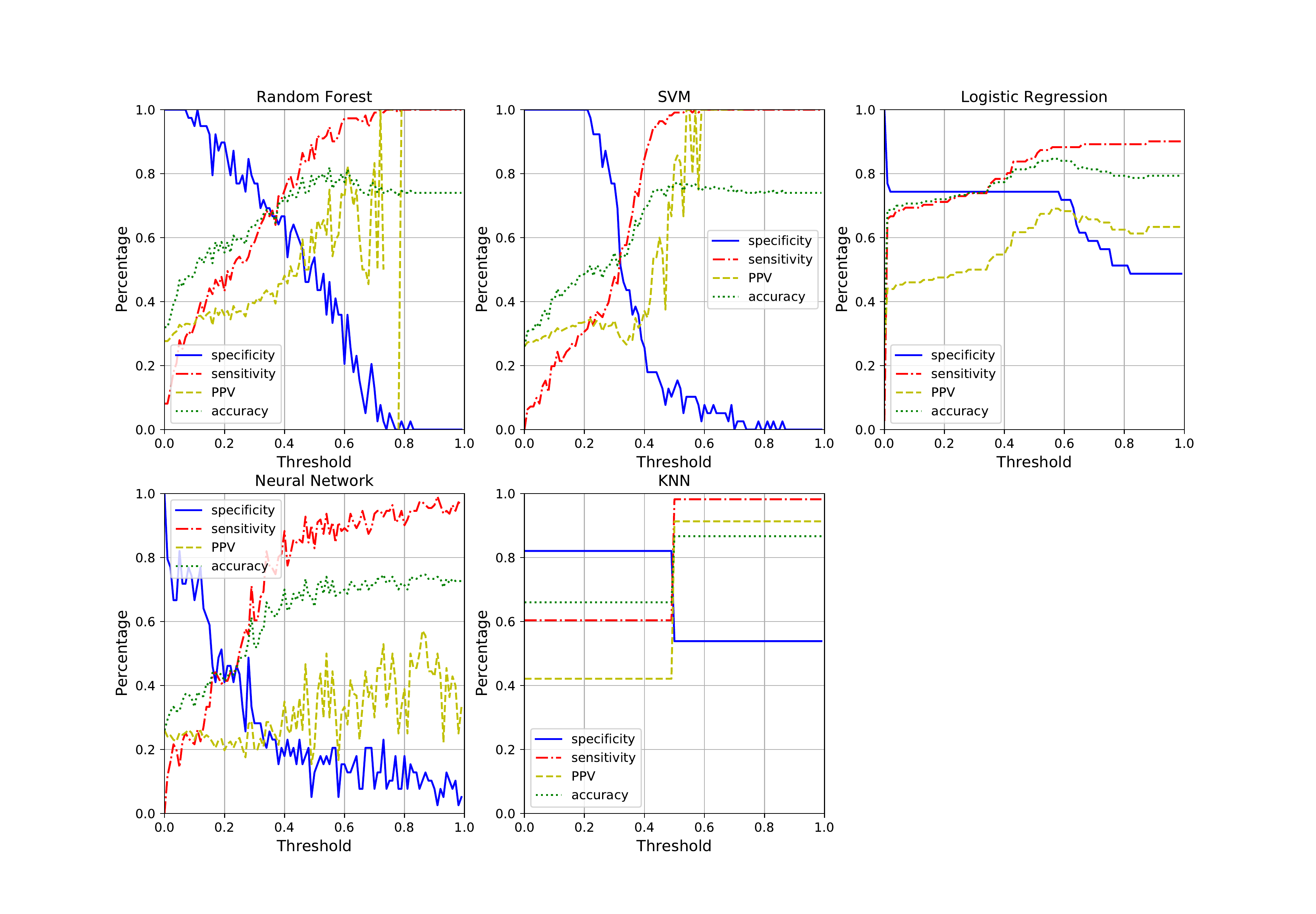}
        \caption{Sensitivity, specificity, PPV and accuracy varying along with threshold. The appropriate threshold is chosen according to the specific purpose.}\label{fig:thres}
\end{figure}

In the experiment, we optimize the model by performing grid search to find the best hyperparameter setting. We compare the model performance in terms of specificity, PPV and accuracy by fixing the sensitivity to around $0.95$, because our goal is to correctly identify the risk patients and have low false alarm rate at the same time. We also evaluate the area under the ROC curve (AUC-ROC) and area under the precision-recall curve (AUC-PR). Since our training class is a skew distribution and the number of negative examples greatly exceeds the number of positives examples (18 vs. 7), AUC-PR is very useful to compare false positives with true positives. Table~\ref{tab:eval} summarizes the evaluation results from all models with sensitivity fixed to around $0.95$. From the results, we observe that KNN has the highest AUC-PR, Specificity and PPV value and second highest AUC-ROC. The results indicate that KNN can achieve very high sensitivity with acceptable false alarm rate compared with other models. Neural network does not seem to work in our task, the possible reason could be overfitting. Figure \ref{fig:thres} shows the specificity, sensitivity PPV and accuracy varying with different threshold. The figure of performance metrics with varying threshold can help us tune sensitivity, specificity, PPV and accuracy according to specific purposes. For many clinical applications it is desirable to achieve high sensitivity as well as high specificity, which is able to avoid generating false alarms.

\begin{figure}
\begin{minipage}[c]{0.39\linewidth}
		\includegraphics[width=\linewidth]{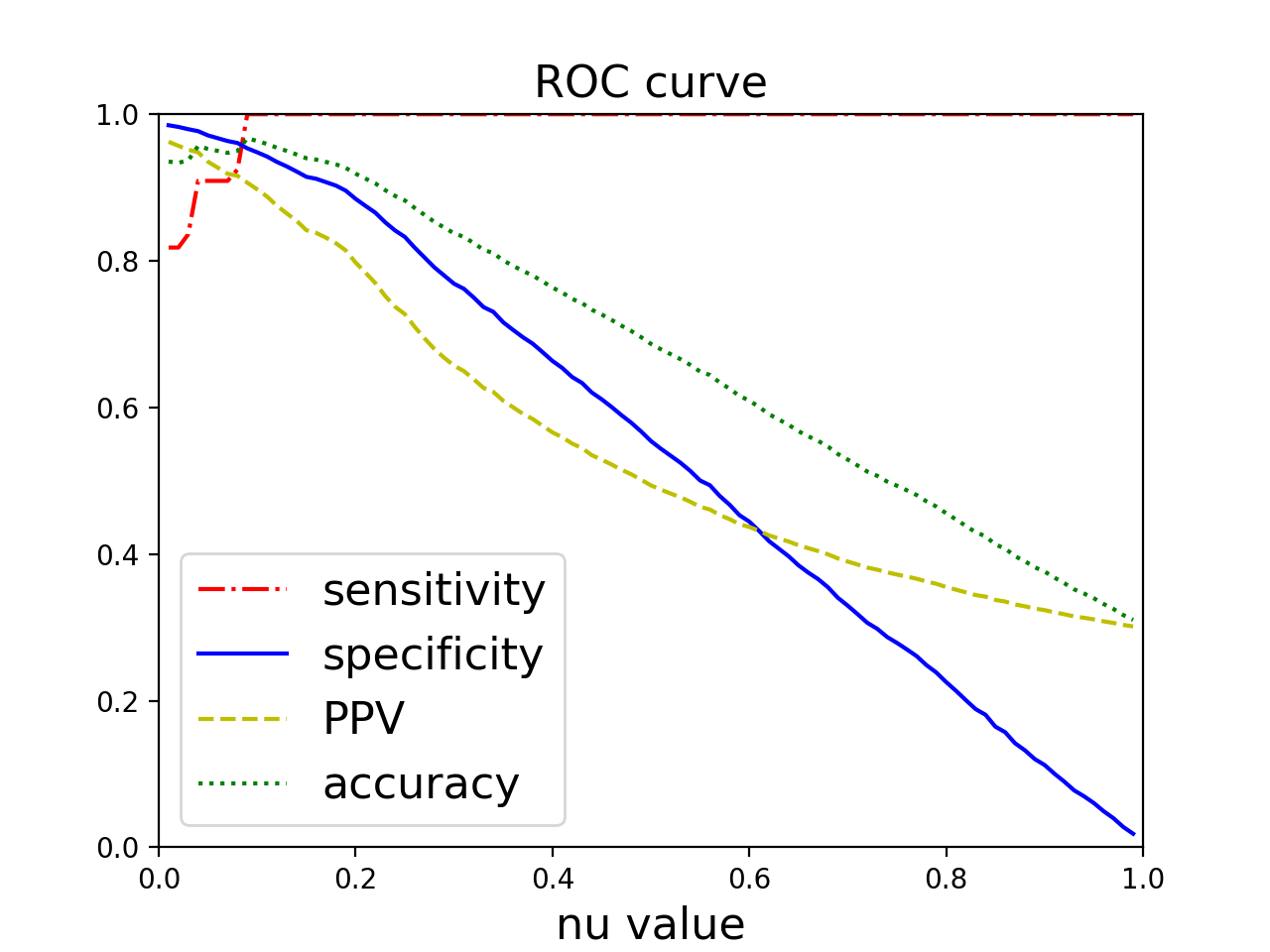}
        \caption{Sensitivity, specificity, PPV and accuracy vary along with $\nu$. The best $\nu$ is 0.9 for our dataset.}\label{fig:nu_roc}
\end{minipage}
\begin{minipage}[c]{0.39\linewidth}
		\includegraphics[width=\linewidth]{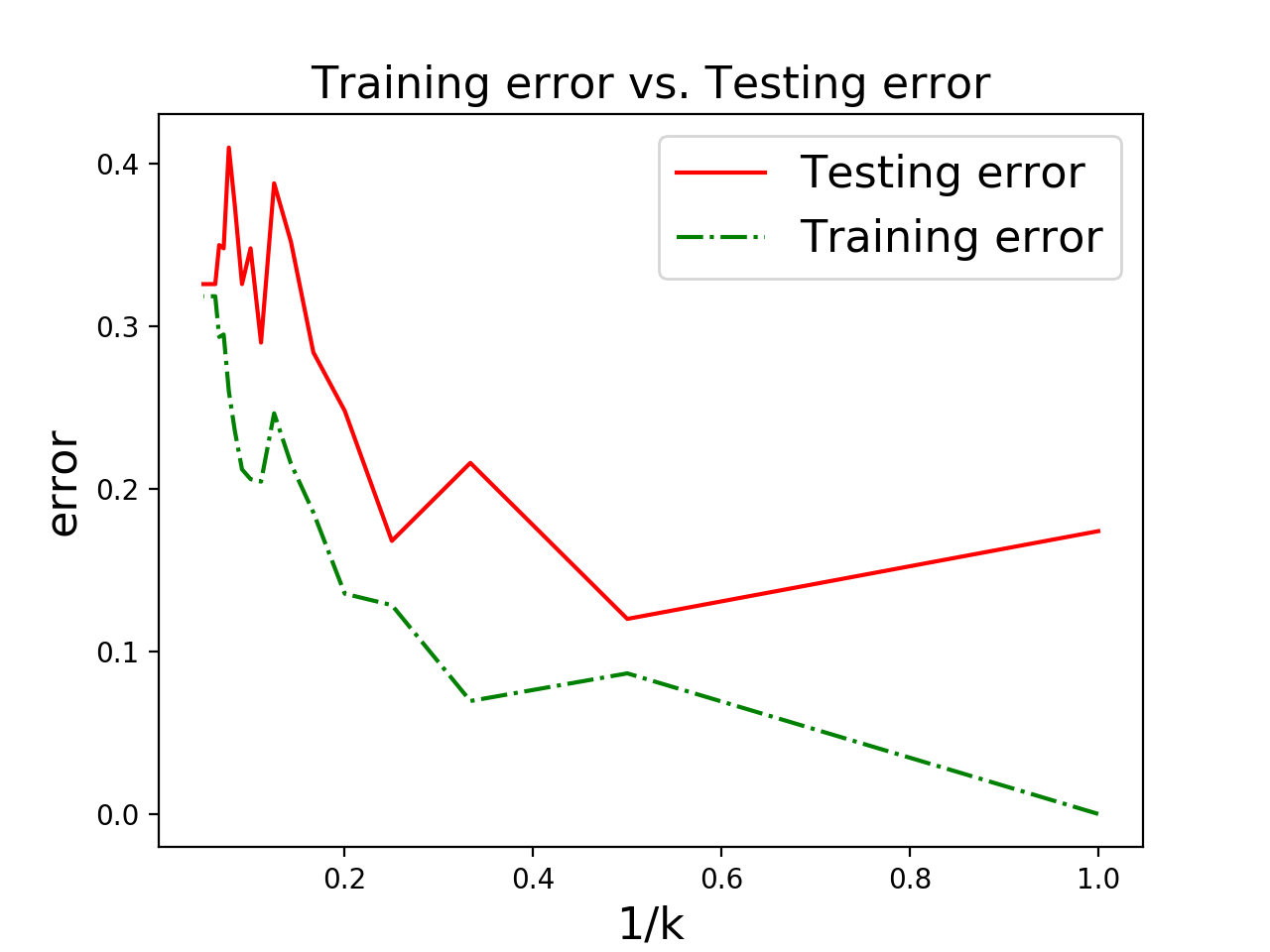}
        \caption{Generalization performance with the inverse of number of nearest neighbors}\label{fig:generalization_plot}
\end{minipage}
\end{figure}

However, for all models, overfitting could probably occur when the training set is small. A commonly used justification for overfitting is to compare training error with testing error. If the testing error is much higher than the training error, the model is very likely overfitting. On the other hand, if the testing error is very close to the training error, it implies the model generalizes well, thus dose not overfit. The results of performance evaluation shown on Table~\ref{tab:eval} suggest KNN is the best model for our case. Hence, we focus on fine-tuning the parameters of KNN to make it generalize better. Figure~\ref{fig:generalization_plot} shows the influence of the number of nearest neighbor on the model's generalization ability. Moreover, as suggested by Occam's razor principle in machine learning, smaller model with the same accuracy is preferred over the complex model. Hence, we apply sequential forward feature selection to select the smallest feature subset when there are multiple subsets leading to same accuracy. The degree of overfitting is the smallest when $K=2$, where the model can achieve the accuracy of $0.880$ for the unseen data. The KNN with 2 nearest neighbors is used throughout the experiment, including the later part of the section where we evaluate the impacts of multiple data modalities and how early we can predict the deterioration. 

We also evaluate the accuracy of using the LACE index, shown on Figure~\ref{tab:eval}. LACE index is calculated by incorporating four parameters. "L" stands for length of stay in the hospital. "A" stands for acuity of admission of patient in the hospital. "C" stands for co-morbidity. "E" stands for number of emergency department visit. In our experiment, the threshold of LACE index is 10 (as suggested by \cite{Wang2014,Low2015,ElMorr2017}), which means the patients will be predicted as readmitted if their LACE index are larger than $10$. The KNN model outperforms the LACE index in terms of sensitivity, PPV and overall accuracy, which demonstrates the feasibility of predicting the deterioration risk of patient via the data passively collected by Fitbit. \\

\begin{table} [h]
\begin{tabular} { | c | c | c | c | c | c | c |}
\hline
 Model & AUC-ROC & AUC-PR &  Specificity & Sensitivity & PPV & Accuracy \\ \hline
 Random Forest & 0.7434 & 0.5551 & 0.3077 & 0.9459 & 0.6667 & 0.780
 \\ \hline
 SVM & 0.5943 & 0.4016 & 0.1795 & 0.9459 & 0.5385 & 0.7467
 \\ \hline
 Logistic Regression & 0.7829 & 0.6118 & 0.4872 & 0.9009 & 0.6333 & 0.7933
 \\ \hline
 Neural Network & 0.4002 & 0.2048 & 0.077 & 0.9459 & 0.3333 & 0.720
 \\ \hline
 KNN & 0.7533 & \textbf{0.6880} & \textbf{0.5385} & 0.9820 & \textbf{0.9130} & \textbf{0.8667}
 \\ \hline
 LACE & & & 0.7647 & 0.6250 & 0.5556 & 0.7826 \\
 \hline
\end{tabular}
\caption {Performance comparison of different models for predicting deterioration. The results shown on the table are based on fixing sensitivity to be around 0.95. In general, KNN is a good model for predicting risk of deterioration.} \label{tab:eval} 
\end{table}
\begin{table} [h!]
\begin{tabular}{|c|c|c|c|c|}
\hline
Random forest & SVM & Logistic regression & Neural network & KNN \\ 
\hline
\parbox[t]{3cm}{min restless count \\ avg minasleep \\ DFA sleep 360 minutes \\ daily sedentary bout \\ HR Local Homogeneity \\} & \parbox[t]{3cm}{avg minasleep \\ DFA sleep 360 minutes \\  min minute awake \\} & \parbox[t]{3cm}{avg time in bed \\ avg minute awake \\ avg awake count \\ avg restless count \\ max restless duration \\ avg daily step \\ activity quality \\ DFA HR 10 minutes \\ DFA sleep 60 minutes \\ DFA sleep 120 minutes \\} & \parbox[t]{3cm}{avg awake count \\ avg restless count \\ min minute awake \\ max awake count \\ min daily step \\ avg daily step \\ DFA sleep 360 minutes \\ } & \parbox[t]{3cm}{all features \textbf{except}: \\ HR Local Homogeneity \\ sleep efficiency \\ sedentary per bout \\ } \\
\hline
\end{tabular}
\caption{Features selected by Sequential Forward Feature Selection} \label{tab:ft1}
\end{table}
\subsubsection{Impacts of Multiple Data Modalities}
We now evaluate the contributions of different data modalities to prediction. The features used in our machine learning models are derived from three modalities collected by the Fitbit Charge HR: heart rate, step and sleep. In the following we compare the accuracy of the KNN models when trained with different combinations of data modalities. 
\begin{table} [h!]
\begin{tabular} { | c | c | c | c | c | c | c | c |}
\hline
 Model & Step & HR & Sleep & Sleep, HR & Sleep, Step & Step, HR & All\\ \hline
 Best Accuracy & 0.7960 & 0.832 & 0.728 & 0.820 & 0.860 & 0.852 & 0.880 \\ \hline
\end{tabular}
\caption {Best accuracy of models trained with different type of features, indicating all features have contribution to overall accuracy} \label{tab:feature_loo} 
\end{table}

\begin{table} [h!]
\begin{tabular} { | c | c | c | c | c | c | c | c |}
\hline
 Model & AUC-ROC & AUC-PR &  Specificity & Sensitivity & PPV & Accuracy\\ \hline
 Step & 0.6910 & 0.5839 & 0.3333 & 0.9550 & 0.7222 & 0.7933 \\ \hline
 HR & 0.8064 & 0.6502 & 0.3590 & 0.9369 & 0.6667 & 0.7867 \\ \hline
 Sleep & 0.7237 & 0.3439 & 0.0513 & 0.9099 & 0.1667 & 0.6867 \\ \hline
 Sleep, HR & 0.8064 & 0.6502 & 0.3590 & 0.9369 & 0.6667 & 0.7867 \\ \hline
 Sleep, Step & 0.6556 & 0.6304 & 0.4615 & 1.0 & 1.0 & 0.860 \\ \hline
 Step, HR & 0.8151 & 0.6991 & 0.4872 & 0.9369 & 0.7308 & 0.820 \\ \hline
 All & 0.7533 & 0.6880 & 0.5385 & 0.9820 & 0.9130 & 0.8667 \\ \hline
\end{tabular}
\caption {Performance metrics of models trained with different combination of features. The sensitivity is fixed to be around 0.95. Better performance is achieved using features from multiple sensing modalities.} \label{tab:feature_fixed} 
\end{table}

In a state-of-the-art study \cite{Bae2016}, Bae et al. trained random forest using only features derived from step. In comparison, our results show that incorporating multi-modal data in machine learning models can further improve the prediction accuracy, as shown on Table~\ref{tab:feature_loo}. For instance, the best accuracy of KNN model trained with a combination of heart rate, sleep and step features is $0.880$, compared with $0.7280$ when using features derived from sleep only. Table~\ref{tab:feature_fixed} summarizes the performance metrics for all models with sensitivity fixed as $0.95$. In general, using all the features produces better results from all metrics. Moreover, we find that the model trained with sleep and step features can also give good performance. Overall, combining features derived from step, heart rate and sleep consistently improve model accuracy. This result motivates combining multi-modal data for predicting clinical outcomes.

\subsubsection{How early can we predict clinical deterioration?}

We investigate the ability of the model in predicting the deterioration ahead of time. In order to simulate this situation, we train and evaluate the models using 5-day, 10-day, 15-day and 20-day data from the beginning of the monitoring. As shown on Table~\ref{tab:difdays}, the prediction accuracy increases as using longer period of measurement. As shown on Table~\ref{tab:day_fixed}, the specificity increases drastically as using more days when fixing the sensitivity to be near $0.95$. In particular, 20-day data yield the highest overall accuracy $0.8677$ as well as highest specificity $0.5385$. The results suggest that long-term monitoring data should be used for predicting the risk of deterioration.


\begin{table} [h!]
\begin{tabular} { | c | c | c | c | c | }
\hline
 Length of day & 5-day & 10-day & 15-day & 20-day\\ \hline
Best Accuracy & 0.7340 & 0.8420 & 0.8280 & 0.880 \\ \hline
\end{tabular}
\caption {Accuracy of models trained with different length of day. The accuracy increases as using more days of data for prediction.} \label{tab:difdays} 
\end{table}
\begin{table} [h!]
\begin{tabular} { | c | c | c | c | c | c | c | c |}
\hline
 Model & AUC-ROC & AUC-PR &  Specificity & Sensitivity & PPV & Accuracy\\ \hline
 5-day & 0.4247 & 0.2114 & 0.0256 & 0.9009 & 0.0833 & 0.6733 \\ \hline
 10-day & 0.7339 & 0.6884 & 0.3077 & 1.0 & 1.0 & 0.820 \\ \hline
 15-day & 0.7710 & 0.5808 & 0.3333 & 0.9009 & 0.5417 & 0.7533 \\ \hline
 20-day & 0.7533 & 0.6880 & 0.5385 & 0.9820 & 0.9130 & 0.8667 \\ \hline
\end{tabular}
\caption {Performance metrics for models trained with different length of day. The sensitivity is fixed to be around 0.95.} \label{tab:day_fixed} 
\end{table}
\section{Conclusion}
In this paper, we explore the feasibility and potential of using wearables to predict clinical deterioration among outpatients through a clinical study involving 25 heart failure patients discharged from a hospital. Our primary findings from the study were two fold. First, our experience demonstrated the feasibility of collecting multi-modal data (step, sleep and heart rate) from outpatients using wristbands. 88\% of participants wore the Fitbit Charge HR wristband regularly which indicated a high level of \textit{compliance}. The monitoring system achieved high data \textit{yield}, collecting 80\% of per-minute step data from 84\% of the participants. While the heart rate data yield was lower, the median gap between data samples were 4 minutes. Furthermore, the monitoring system collected 73\% of the data from the wristband to our cloud-based database within an hour. 

Second, We demonstrate the potential of machine learning models to predict clinical deterioration among outpatients. Our proposed weighted samples one class SVM can reach 0.9635 accuracy for deterioration early warning. Our results showed machine learning models can exploit multi-modal data to achieve high accuracy for identifying patient's risk of deterioration (e.g., 0.88 under K nearest neighbor vs. 0.78 with the LACE index). Moreover, machine learning models can predict deterioration risk using only the first ten days of data collected, which suggests the potential for early identifying the high risk patients and allows timely intervention. 

Building upon the promising results, we plan to conduct larger clinical studies to generalize the results and impact clinical practice. 

\bibliographystyle{ACM-Reference-Format}
\bibliography{sample-bibliography}

%% file: sample-acmlarge.bbl

\begin{thebibliography}{40}


\ifx \showCODEN    \undefined \def \showCODEN     #1{\unskip}     \fi
\ifx \showDOI      \undefined \def \showDOI       #1{#1}\fi
\ifx \showISBNx    \undefined \def \showISBNx     #1{\unskip}     \fi
\ifx \showISBNxiii \undefined \def \showISBNxiii  #1{\unskip}     \fi
\ifx \showISSN     \undefined \def \showISSN      #1{\unskip}     \fi
\ifx \showLCCN     \undefined \def \showLCCN      #1{\unskip}     \fi
\ifx \shownote     \undefined \def \shownote      #1{#1}          \fi
\ifx \showarticletitle \undefined \def \showarticletitle #1{#1}   \fi
\ifx \showURL      \undefined \def \showURL       {\relax}        \fi
\providecommand\bibfield[2]{#2}
\providecommand\bibinfo[2]{#2}
\providecommand\natexlab[1]{#1}
\providecommand\showeprint[2][]{arXiv:#2}

\bibitem[\protect\citeauthoryear{Abdulmajeed}{Abdulmajeed}{2016}]%
        {Abdulmajeed16}
\bibfield{author}{\bibinfo{person}{Raghad Abdulmajeed}.}
  \bibinfo{year}{2016}\natexlab{}.
\newblock {\em \bibinfo{title}{The Use of Continuous Monitoring of Heart Rate
  as a Prognosticator of Readmission in Heart Failure Patients}}.
\newblock \bibinfo{thesistype}{Master's\ thesis}. \bibinfo{school}{University
  of Toronto}, \bibinfo{address}{Canada}.
\newblock


\bibitem[\protect\citeauthoryear{Amer, Goldstein, and Abdennadher}{Amer
  et~al\mbox{.}}{2013}]%
        {Amer2013}
\bibfield{author}{\bibinfo{person}{Mennatallah Amer}, \bibinfo{person}{Markus
  Goldstein}, {and} \bibinfo{person}{Slim Abdennadher}.}
  \bibinfo{year}{2013}\natexlab{}.
\newblock \showarticletitle{Enhancing One-class Support Vector Machines for
  Unsupervised Anomaly Detection}. In \bibinfo{booktitle}{{\em Proceedings of
  the ACM SIGKDD Workshop on Outlier Detection and Description}} {\em
  (\bibinfo{series}{ODD '13})}. \bibinfo{publisher}{ACM}, \bibinfo{address}{New
  York, NY, USA}, \bibinfo{pages}{8--15}.
\newblock
\showISBNx{978-1-4503-2335-2}
\showDOI{%
\url{https://doi.org/10.1145/2500853.2500857}}


\bibitem[\protect\citeauthoryear{Andrews~RM}{Andrews~RM}{2006}]%
        {andrews07}
\bibfield{author}{\bibinfo{person}{Elixhauser~A. Andrews~RM}.}
  \bibinfo{year}{2006}\natexlab{}.
\newblock \showarticletitle{The National Hospital Bill: Growth Trends and 2005
  Update on the Most Expensive Conditions by Payer}.
\newblock \bibinfo{journal}{{\em Healthcare Cost and Utilization Project (HCUP)
  Statistical Briefs [Internet]\/}} (\bibinfo{year}{2006}).
\newblock
\showURL{%
\url{https://www.ncbi.nlm.nih.gov/books/NBK56314/}}


\bibitem[\protect\citeauthoryear{Appelboom, Taylor, Bruce, Bassile, Malakidis,
  Yang, Youngerman, D'Amico, Bruce, Bruy{\`e}re, Reginster, Dumont, and
  Connolly~Jr}{Appelboom et~al\mbox{.}}{2015}]%
        {Appelboom15}
\bibfield{author}{\bibinfo{person}{Geoff Appelboom}, \bibinfo{person}{E.~Blake
  Taylor}, \bibinfo{person}{Eliza Bruce}, \bibinfo{person}{C.~Clare Bassile},
  \bibinfo{person}{Corinna Malakidis}, \bibinfo{person}{Annie Yang},
  \bibinfo{person}{Brett Youngerman}, \bibinfo{person}{Randy D'Amico},
  \bibinfo{person}{Sam Bruce}, \bibinfo{person}{Olivier Bruy{\`e}re},
  \bibinfo{person}{Jean-Yves Reginster}, \bibinfo{person}{PL~Emmanuel Dumont},
  {and} \bibinfo{person}{Sander~E. Connolly~Jr}.}
  \bibinfo{year}{2015}\natexlab{}.
\newblock \showarticletitle{Mobile Phone-Connected Wearable Motion Sensors to
  Assess Postoperative Mobilization}.
\newblock \bibinfo{journal}{{\em JMIR mHealth uHealth\/}} \bibinfo{volume}{3},
  \bibinfo{number}{3} (\bibinfo{date}{28 Jul} \bibinfo{year}{2015}),
  \bibinfo{pages}{e78}.
\newblock
\showDOI{%
\url{https://doi.org/10.2196/mhealth.3785}}


\bibitem[\protect\citeauthoryear{B and GC}{B and GC}{2015}]%
        {Ziaeian15}
\bibfield{author}{\bibinfo{person}{Ziaeian B} {and} \bibinfo{person}{Fonarow
  GC}.} \bibinfo{year}{2015}\natexlab{}.
\newblock \showarticletitle{The Prevention of Hospital Readmissions in Heart
  Failure}.
\newblock \bibinfo{journal}{{\em Progress in Cardiovascular Diseases\/}}
  \bibinfo{volume}{58}, \bibinfo{number}{4} (\bibinfo{year}{2015}),
  \bibinfo{pages}{379--385}.
\newblock
\showDOI{%
\url{https://doi.org/10.1016/j.pcad.2015.09.004}}


\bibitem[\protect\citeauthoryear{Bae, Dey, and Low}{Bae et~al\mbox{.}}{2016}]%
        {Bae2016}
\bibfield{author}{\bibinfo{person}{Sangwon Bae}, \bibinfo{person}{Anind~K.
  Dey}, {and} \bibinfo{person}{Carissa~A. Low}.}
  \bibinfo{year}{2016}\natexlab{}.
\newblock \showarticletitle{Using Passively Collected Sedentary Behavior to
  Predict Hospital Readmission}. In \bibinfo{booktitle}{{\em Proceedings of the
  2016 ACM International Joint Conference on Pervasive and Ubiquitous
  Computing}} {\em (\bibinfo{series}{UbiComp '16})}. \bibinfo{publisher}{ACM},
  \bibinfo{address}{New York, NY, USA}, \bibinfo{pages}{616--621}.
\newblock
\showISBNx{978-1-4503-4461-6}
\showDOI{%
\url{https://doi.org/10.1145/2971648.2971750}}


\bibitem[\protect\citeauthoryear{Beleites and Salzer}{Beleites and
  Salzer}{2008}]%
        {Beleites2008}
\bibfield{author}{\bibinfo{person}{Claudia Beleites} {and}
  \bibinfo{person}{Reiner Salzer}.} \bibinfo{year}{2008}\natexlab{}.
\newblock \showarticletitle{Assessing and improving the stability of
  chemometric models in small sample size situations}.
\newblock \bibinfo{journal}{{\em Analytical and Bioanalytical Chemistry\/}}
  \bibinfo{volume}{390}, \bibinfo{number}{5} (\bibinfo{date}{01 Mar}
  \bibinfo{year}{2008}), \bibinfo{pages}{1261--1271}.
\newblock
\showDOI{%
\url{https://doi.org/10.1007/s00216-007-1818-6}}


\bibitem[\protect\citeauthoryear{Cadmus-Bertram, Marcus, Patterson, Parker, and
  Morey}{Cadmus-Bertram et~al\mbox{.}}{2015}]%
        {cadmus15}
\bibfield{author}{\bibinfo{person}{Lisa Cadmus-Bertram},
  \bibinfo{person}{H.~Bess Marcus}, \bibinfo{person}{E.~Ruth Patterson},
  \bibinfo{person}{A.~Barbara Parker}, {and} \bibinfo{person}{L.~Brittany
  Morey}.} \bibinfo{year}{2015}\natexlab{}.
\newblock \showarticletitle{Use of the Fitbit to Measure Adherence to a
  Physical Activity Intervention Among Overweight or Obese, Postmenopausal
  Women: Self-Monitoring Trajectory During 16 Weeks}.
\newblock \bibinfo{journal}{{\em JMIR mHealth uHealth\/}} \bibinfo{volume}{3},
  \bibinfo{number}{4} (\bibinfo{date}{19 Nov} \bibinfo{year}{2015}),
  \bibinfo{pages}{e96}.
\newblock
\showDOI{%
\url{https://doi.org/10.2196/mhealth.4229}}


\bibitem[\protect\citeauthoryear{Chipara, Lu, Bailey, and Roman}{Chipara
  et~al\mbox{.}}{2010}]%
        {Chipara2010}
\bibfield{author}{\bibinfo{person}{Octav Chipara}, \bibinfo{person}{Chenyang
  Lu}, \bibinfo{person}{Thomas~C. Bailey}, {and} \bibinfo{person}{Gruia-Catalin
  Roman}.} \bibinfo{year}{2010}\natexlab{}.
\newblock \showarticletitle{Reliable Clinical Monitoring Using Wireless Sensor
  Networks: Experiences in a Step-down Hospital Unit}. In
  \bibinfo{booktitle}{{\em Proceedings of the 8th ACM Conference on Embedded
  Networked Sensor Systems}} {\em (\bibinfo{series}{SenSys '10})}.
  \bibinfo{publisher}{ACM}, \bibinfo{address}{New York, NY, USA},
  \bibinfo{pages}{155--168}.
\newblock
\showISBNx{978-1-4503-0344-6}
\showDOI{%
\url{https://doi.org/10.1145/1869983.1869999}}


\bibitem[\protect\citeauthoryear{Cotter, Bhalla, Wallis, and Biram}{Cotter
  et~al\mbox{.}}{2012}]%
        {Cotter2012}
\bibfield{author}{\bibinfo{person}{Paul~E. Cotter}, \bibinfo{person}{Vikas~K.
  Bhalla}, \bibinfo{person}{Stephen~J. Wallis}, {and} \bibinfo{person}{Richard
  W.~S. Biram}.} \bibinfo{year}{2012}\natexlab{}.
\newblock \showarticletitle{Predicting readmissions: poor performance of the
  LACE index in an older UK population}.
\newblock \bibinfo{journal}{{\em Age and Ageing\/}} \bibinfo{volume}{41},
  \bibinfo{number}{6} (\bibinfo{year}{2012}), \bibinfo{pages}{784--789}.
\newblock
\showDOI{%
\url{https://doi.org/10.1093/ageing/afs073}}


\bibitem[\protect\citeauthoryear{D, H, A, and et~al}{D et~al\mbox{.}}{2011}]%
        {Kansagara11}
\bibfield{author}{\bibinfo{person}{Kansagara D}, \bibinfo{person}{Englander H},
  \bibinfo{person}{Salanitro A}, {and} \bibinfo{person}{et al}.}
  \bibinfo{year}{2011}\natexlab{}.
\newblock \showarticletitle{Risk prediction models for hospital readmission: A
  systematic review}.
\newblock \bibinfo{journal}{{\em JAMA\/}} \bibinfo{volume}{306},
  \bibinfo{number}{15} (\bibinfo{year}{2011}), \bibinfo{pages}{1688--1698}.
\newblock
\showDOI{%
\url{https://doi.org/10.1001/jama.2011.1515}}


\bibitem[\protect\citeauthoryear{de~Zambotti, Goldstone, Claudatos, Colrain,
  and Baker}{de~Zambotti et~al\mbox{.}}{2017}]%
        {Zambotti17}
\bibfield{author}{\bibinfo{person}{Massimiliano de Zambotti},
  \bibinfo{person}{Aimee Goldstone}, \bibinfo{person}{Stephanie Claudatos},
  \bibinfo{person}{Ian~M. Colrain}, {and} \bibinfo{person}{Fiona~C. Baker}.}
  \bibinfo{year}{2017}\natexlab{}.
\newblock \showarticletitle{A validation study of Fitbit Charge 2 compared with
  polysomnography in adults}.
\newblock \bibinfo{journal}{{\em Chronobiology International\/}}
  \bibinfo{volume}{0}, \bibinfo{number}{0} (\bibinfo{year}{2017}),
  \bibinfo{pages}{1--12}.
\newblock
\showDOI{%
\url{https://doi.org/10.1080/07420528.2017.1413578}}
\showeprint{https://doi.org/10.1080/07420528.2017.1413578}
\newblock
\shownote{PMID: 29235907.}


\bibitem[\protect\citeauthoryear{Donz{\'e}, Williams, Robinson, Zimlichman,
  Aujesky, Vasilevskis, Kripalani, Metlay, Wallington, Fletcher, Auerbach, and
  Schnipper}{Donz{\'e} et~al\mbox{.}}{2016}]%
        {Donzé2016}
\bibfield{author}{\bibinfo{person}{Jacques~D. Donz{\'e}},
  \bibinfo{person}{Mark~V. Williams}, \bibinfo{person}{Edmondo~J. Robinson},
  \bibinfo{person}{Eyal Zimlichman}, \bibinfo{person}{Drahomir Aujesky},
  \bibinfo{person}{Eduard~E. Vasilevskis}, \bibinfo{person}{Sunil Kripalani},
  \bibinfo{person}{Joshua~P. Metlay}, \bibinfo{person}{Tamara Wallington},
  \bibinfo{person}{Grant~S. Fletcher}, \bibinfo{person}{Andrew~D. Auerbach},
  {and} \bibinfo{person}{Jeffrey~L. Schnipper}.}
  \bibinfo{year}{2016}\natexlab{}.
\newblock \showarticletitle{International Validity of the "HOSPITAL" Score to
  Predict 30-day Potentially Avoidable Readmissions in Medical Patients}.
\newblock \bibinfo{journal}{{\em JAMA Intern Med\/}} \bibinfo{volume}{176},
  \bibinfo{number}{4} (\bibinfo{date}{Apr} \bibinfo{year}{2016}),
  \bibinfo{pages}{496--502}.
\newblock
\showISSN{2168-6106}
\showDOI{%
\url{https://doi.org/10.1001/jamainternmed.2015.8462}}
\newblock
\shownote{26954698[pmid].}


\bibitem[\protect\citeauthoryear{E}{E}{2013}]%
        {marks13}
\bibfield{author}{\bibinfo{person}{Marks E}.} \bibinfo{year}{2013}\natexlab{}.
\newblock \showarticletitle{Complexity science and the readmission dilemma:
  Comment on "potentially avoidable 30-day hospital readmissions in medical
  patients" and "association of self-reported hospital discharge handoffs with
  30-day readmissions"}.
\newblock \bibinfo{journal}{{\em JAMA Internal Medicine\/}}
  \bibinfo{volume}{173}, \bibinfo{number}{8} (\bibinfo{year}{2013}),
  \bibinfo{pages}{629--631}.
\newblock
\showDOI{%
\url{https://doi.org/10.1001/jamainternmed.2013.4065}}


\bibitem[\protect\citeauthoryear{El~Morr, Ginsburg, Nam, and Woollard}{El~Morr
  et~al\mbox{.}}{2017}]%
        {ElMorr2017}
\bibfield{author}{\bibinfo{person}{Christo El~Morr}, \bibinfo{person}{Liane
  Ginsburg}, \bibinfo{person}{Seungree Nam}, {and} \bibinfo{person}{Susan
  Woollard}.} \bibinfo{year}{2017}\natexlab{}.
\newblock \showarticletitle{Assessing the Performance of a Modified LACE Index
  (LACE-rt) to Predict Unplanned Readmission After Discharge in a Community
  Teaching Hospital}.
\newblock \bibinfo{journal}{{\em Interact J Med Res\/}} \bibinfo{volume}{6},
  \bibinfo{number}{1} (\bibinfo{date}{08 Mar} \bibinfo{year}{2017}),
  \bibinfo{pages}{e2}.
\newblock
\showISSN{1929-073X}
\showDOI{%
\url{https://doi.org/10.2196/ijmr.7183}}
\newblock
\shownote{v6i1e2[PII].}


\bibitem[\protect\citeauthoryear{Fisher, Kuo, Sharma, Raji, Kumar, Goodwin,
  Ostir, and Ottenbacher}{Fisher et~al\mbox{.}}{2013}]%
        {Fisher2013}
\bibfield{author}{\bibinfo{person}{Steve~R. Fisher}, \bibinfo{person}{Yong-Fang
  Kuo}, \bibinfo{person}{Gulshan Sharma}, \bibinfo{person}{Mukaila~A. Raji},
  \bibinfo{person}{Amit Kumar}, \bibinfo{person}{James~S. Goodwin},
  \bibinfo{person}{Glenn~V. Ostir}, {and} \bibinfo{person}{Kenneth~J.
  Ottenbacher}.} \bibinfo{year}{2013}\natexlab{}.
\newblock \showarticletitle{Mobility After Hospital Discharge as a Marker for
  30-Day Readmission}.
\newblock \bibinfo{journal}{{\em J Gerontol A Biol Sci Med Sci\/}}
  \bibinfo{volume}{68}, \bibinfo{number}{7} (\bibinfo{date}{19 Jul}
  \bibinfo{year}{2013}), \bibinfo{pages}{805--810}.
\newblock
\showISSN{1079-5006}
\showDOI{%
\url{https://doi.org/10.1093/gerona/gls252}}
\newblock
\shownote{23254776[pmid].}


\bibitem[\protect\citeauthoryear{He, Mathews, Kalloo, and Hutfless}{He
  et~al\mbox{.}}{2014}]%
        {He2014}
\bibfield{author}{\bibinfo{person}{Danning He}, \bibinfo{person}{Simon~C
  Mathews}, \bibinfo{person}{Anthony~N Kalloo}, {and} \bibinfo{person}{Susan
  Hutfless}.} \bibinfo{year}{2014}\natexlab{}.
\newblock \showarticletitle{Mining high-dimensional administrative claims data
  to predict early hospital readmissions}.
\newblock \bibinfo{journal}{{\em Journal of the American Medical Informatics
  Association\/}} \bibinfo{volume}{21}, \bibinfo{number}{2}
  (\bibinfo{year}{2014}), \bibinfo{pages}{272--279}.
\newblock
\showDOI{%
\url{https://doi.org/10.1136/amiajnl-2013-002151}}


\bibitem[\protect\citeauthoryear{Hosseinzadeh, Izadi, Verma, Precup, and
  Buckeridge}{Hosseinzadeh et~al\mbox{.}}{2013}]%
        {Hosseinzadeh2013}
\bibfield{author}{\bibinfo{person}{Arian Hosseinzadeh},
  \bibinfo{person}{Masoumeh Izadi}, \bibinfo{person}{Aman Verma},
  \bibinfo{person}{Doina Precup}, {and} \bibinfo{person}{David Buckeridge}.}
  \bibinfo{year}{2013}\natexlab{}.
\newblock \showarticletitle{Assessing the Predictability of Hospital
  Readmission Using Machine Learning}. In \bibinfo{booktitle}{{\em Proceedings
  of the Twenty-Seventh AAAI Conference on Artificial Intelligence}} {\em
  (\bibinfo{series}{AAAI'13})}. \bibinfo{publisher}{AAAI Press},
  \bibinfo{pages}{1532--1538}.
\newblock
\showURL{%
\url{http://dl.acm.org/citation.cfm?id=2891460.2891675}}


\bibitem[\protect\citeauthoryear{Hu, Ivanov, Chen, Carpena, and
  Eugene~Stanley}{Hu et~al\mbox{.}}{2001}]%
        {PhysRevE.64.011114}
\bibfield{author}{\bibinfo{person}{Kun Hu}, \bibinfo{person}{Plamen~Ch.
  Ivanov}, \bibinfo{person}{Zhi Chen}, \bibinfo{person}{Pedro Carpena}, {and}
  \bibinfo{person}{H. Eugene~Stanley}.} \bibinfo{year}{2001}\natexlab{}.
\newblock \showarticletitle{Effect of trends on detrended fluctuation
  analysis}.
\newblock \bibinfo{journal}{{\em Phys. Rev. E\/}}  \bibinfo{volume}{64}
  (\bibinfo{date}{Jun} \bibinfo{year}{2001}), \bibinfo{pages}{011114}.
\newblock
Issue 1.
\showDOI{%
\url{https://doi.org/10.1103/PhysRevE.64.011114}}


\bibitem[\protect\citeauthoryear{K, AF, Z, and et~al}{K et~al\mbox{.}}{2013}]%
        {Dharmarajan13}
\bibfield{author}{\bibinfo{person}{Dharmarajan K}, \bibinfo{person}{Hsieh AF},
  \bibinfo{person}{Lin Z}, {and} \bibinfo{person}{et al}.}
  \bibinfo{year}{2013}\natexlab{}.
\newblock \showarticletitle{Diagnoses and timing of 30-day readmissions after
  hospitalization for heart failure, acute myocardial infarction, or
  pneumonia}.
\newblock \bibinfo{journal}{{\em JAMA Internal Medicine\/}}
  \bibinfo{volume}{309}, \bibinfo{number}{4} (\bibinfo{year}{2013}),
  \bibinfo{pages}{355--363}.
\newblock
\showDOI{%
\url{https://doi.org/10.1001/jama.2012.216476}}


\bibitem[\protect\citeauthoryear{Kaewkannate and Kim}{Kaewkannate and
  Kim}{2016}]%
        {Kaewkannate2016}
\bibfield{author}{\bibinfo{person}{Kanitthika Kaewkannate} {and}
  \bibinfo{person}{Soochan Kim}.} \bibinfo{year}{2016}\natexlab{}.
\newblock \showarticletitle{A comparison of wearable fitness devices}.
\newblock \bibinfo{journal}{{\em BMC Public Health\/}} \bibinfo{volume}{16},
  \bibinfo{number}{1} (\bibinfo{date}{24 May} \bibinfo{year}{2016}),
  \bibinfo{pages}{433}.
\newblock
\showISSN{1471-2458}
\showDOI{%
\url{https://doi.org/10.1186/s12889-016-3059-0}}


\bibitem[\protect\citeauthoryear{Low, Lee, Hock~Ong, Wang, Tan, Thumboo, and
  Liu}{Low et~al\mbox{.}}{2015}]%
        {Low2015}
\bibfield{author}{\bibinfo{person}{Lian~Leng Low}, \bibinfo{person}{Kheng~Hock
  Lee}, \bibinfo{person}{Marcus~Eng Hock~Ong}, \bibinfo{person}{Sijia Wang},
  \bibinfo{person}{Shu~Yun Tan}, \bibinfo{person}{Julian Thumboo}, {and}
  \bibinfo{person}{Nan Liu}.} \bibinfo{year}{2015}\natexlab{}.
\newblock \showarticletitle{Predicting 30-Day Readmissions: Performance of the
  LACE Index Compared with a Regression Model among General Medicine Patients
  in Singapore}.
\newblock \bibinfo{journal}{{\em Biomed Res Int\/}}  \bibinfo{volume}{2015}
  (\bibinfo{date}{23 Nov} \bibinfo{year}{2015}), \bibinfo{pages}{169870}.
\newblock
\showISSN{2314-6133}
\showDOI{%
\url{https://doi.org/10.1155/2015/169870}}
\newblock
\shownote{26682212[pmid].}


\bibitem[\protect\citeauthoryear{Mao, Chen, Chen, Lu, Kollef, and Bailey}{Mao
  et~al\mbox{.}}{2012}]%
        {Mao2012}
\bibfield{author}{\bibinfo{person}{Yi Mao}, \bibinfo{person}{Wenlin Chen},
  \bibinfo{person}{Yixin Chen}, \bibinfo{person}{Chenyang Lu},
  \bibinfo{person}{Marin Kollef}, {and} \bibinfo{person}{Thomas Bailey}.}
  \bibinfo{year}{2012}\natexlab{}.
\newblock \showarticletitle{An Integrated Data Mining Approach to Real-time
  Clinical Monitoring and Deterioration Warning}. In \bibinfo{booktitle}{{\em
  Proceedings of the 18th ACM SIGKDD International Conference on Knowledge
  Discovery and Data Mining}} {\em (\bibinfo{series}{KDD '12})}.
  \bibinfo{publisher}{ACM}, \bibinfo{address}{New York, NY, USA},
  \bibinfo{pages}{1140--1148}.
\newblock
\showISBNx{978-1-4503-1462-6}
\showDOI{%
\url{https://doi.org/10.1145/2339530.2339709}}


\bibitem[\protect\citeauthoryear{McDonald, Ledwidge, Cahill, Kelly, Quigley,
  Maurer, Begley, Ryder, Travers, Timmons, and Burke}{McDonald
  et~al\mbox{.}}{2001}]%
        {MaDonald01}
\bibfield{author}{\bibinfo{person}{Kenneth McDonald}, \bibinfo{person}{Mark
  Ledwidge}, \bibinfo{person}{John Cahill}, \bibinfo{person}{Jean Kelly},
  \bibinfo{person}{Peter Quigley}, \bibinfo{person}{Brian Maurer},
  \bibinfo{person}{Fiona Begley}, \bibinfo{person}{Mary Ryder},
  \bibinfo{person}{Bronagh Travers}, \bibinfo{person}{Lorna Timmons}, {and}
  \bibinfo{person}{Teresa Burke}.} \bibinfo{year}{2001}\natexlab{}.
\newblock \showarticletitle{Elimination of early rehospitalization in a
  randomized, controlled trial of multidisciplinary care in a high-risk,
  elderly heart failure population: the potential contributions of specialist
  care, clinical stability and optimal angiotensin-converting enzyme inhibitor
  dose at discharge}.
\newblock \bibinfo{journal}{{\em European Journal of Heart Failure\/}}
  \bibinfo{volume}{3}, \bibinfo{number}{2} (\bibinfo{year}{2001}),
  \bibinfo{pages}{209--215}.
\newblock
\showISSN{1879-0844}
\showDOI{%
\url{https://doi.org/10.1016/S1388-9842(00)00134-3}}


\bibitem[\protect\citeauthoryear{Peng, Buldyrev, Havlin, Simons, Stanley, and
  Goldberger}{Peng et~al\mbox{.}}{1994}]%
        {PhysRevE.49.1685}
\bibfield{author}{\bibinfo{person}{C.-K. Peng}, \bibinfo{person}{S.~V.
  Buldyrev}, \bibinfo{person}{S. Havlin}, \bibinfo{person}{M. Simons},
  \bibinfo{person}{H.~E. Stanley}, {and} \bibinfo{person}{A.~L. Goldberger}.}
  \bibinfo{year}{1994}\natexlab{}.
\newblock \showarticletitle{Mosaic organization of DNA nucleotides}.
\newblock \bibinfo{journal}{{\em Phys. Rev. E\/}}  \bibinfo{volume}{49}
  (\bibinfo{date}{Feb} \bibinfo{year}{1994}), \bibinfo{pages}{1685--1689}.
\newblock
Issue 2.
\showDOI{%
\url{https://doi.org/10.1103/PhysRevE.49.1685}}


\bibitem[\protect\citeauthoryear{Robinson and Hudali}{Robinson and
  Hudali}{2017}]%
        {Robinson2017}
\bibfield{author}{\bibinfo{person}{Robert Robinson} {and}
  \bibinfo{person}{Tamer Hudali}.} \bibinfo{year}{2017}\natexlab{}.
\newblock \showarticletitle{The HOSPITAL score and LACE index as predictors of
  30 day readmission in a retrospective study at a university-affiliated
  community hospital}.
\newblock \bibinfo{journal}{{\em PeerJ\/}}  \bibinfo{volume}{5}
  (\bibinfo{date}{09 Jan} \bibinfo{year}{2017}), \bibinfo{pages}{e3137}.
\newblock
\showISSN{2167-8359}
\showDOI{%
\url{https://doi.org/10.7717/peerj.3137}}
\newblock
\shownote{3137[PII].}


\bibitem[\protect\citeauthoryear{Shameer, Johnson, Yahi, Miotto, Li, Ricks,
  Jebakaran, Kovatch, Sengupta, Gelijns, Moskovitz, Darrow, David, Kasarskis,
  Tatonetti, Pinney, and Dudley}{Shameer et~al\mbox{.}}{[n. d.]}]%
        {SHAMEER2016}
\bibfield{author}{\bibinfo{person}{Khader Shameer}, \bibinfo{person}{Kipp~W
  Johnson}, \bibinfo{person}{Alexandre Yahi}, \bibinfo{person}{Riccardo
  Miotto}, \bibinfo{person}{Li Li}, \bibinfo{person}{Doran Ricks},
  \bibinfo{person}{Jebakumar Jebakaran}, \bibinfo{person}{Patricia Kovatch},
  \bibinfo{person}{Partho~P. Sengupta}, \bibinfo{person}{Sengupta Gelijns},
  \bibinfo{person}{Alan Moskovitz}, \bibinfo{person}{Bruce Darrow},
  \bibinfo{person}{David~L David}, \bibinfo{person}{Andrew Kasarskis},
  \bibinfo{person}{Nicholas~P. Tatonetti}, \bibinfo{person}{Sean Pinney}, {and}
  \bibinfo{person}{Joel~T Dudley}.} \bibinfo{year}{[n. d.]}\natexlab{}.
\newblock \bibinfo{booktitle}{{\em Predictive Modeling of Hospital Readmission
  Rates Using Electronic Medical Record-wide Machine Learning: a Case-study
  Using Mount Sinai Heart Failure Cohort}}.
\newblock \bibinfo{pages}{276--287}.
\newblock
\showDOI{%
\url{https://doi.org/10.1142/9789813207813_0027}}


\bibitem[\protect\citeauthoryear{Shcherbina, Mattsson, Waggott, Salisbury,
  Christle, Hastie, Wheeler, and Ashley}{Shcherbina et~al\mbox{.}}{2017}]%
        {Shcherbina17}
\bibfield{author}{\bibinfo{person}{Anna Shcherbina}, \bibinfo{person}{C.~Mikael
  Mattsson}, \bibinfo{person}{Daryl Waggott}, \bibinfo{person}{Heidi
  Salisbury}, \bibinfo{person}{Jeffrey~W. Christle}, \bibinfo{person}{Trevor
  Hastie}, \bibinfo{person}{Matthew~T. Wheeler}, {and} \bibinfo{person}{Euan~A.
  Ashley}.} \bibinfo{year}{2017}\natexlab{}.
\newblock \showarticletitle{Accuracy in Wrist-Worn, Sensor-Based Measurements
  of Heart Rate and Energy Expenditure in a Diverse Cohort}.
\newblock \bibinfo{journal}{{\em Journal of Personalized Medicine\/}}
  \bibinfo{volume}{7}, \bibinfo{number}{2} (\bibinfo{year}{2017}).
\newblock
\showISSN{2075-4426}
\showDOI{%
\url{https://doi.org/10.3390/jpm7020003}}


\bibitem[\protect\citeauthoryear{Sushmita, Khulbe, Hasan, Newman, Ravindra,
  Roy, Cock, and Teredesai}{Sushmita et~al\mbox{.}}{2016}]%
        {Sushmita2016}
\bibfield{author}{\bibinfo{person}{Shanu Sushmita}, \bibinfo{person}{Garima
  Khulbe}, \bibinfo{person}{Aftab Hasan}, \bibinfo{person}{Stacey Newman},
  \bibinfo{person}{Padmashree Ravindra}, \bibinfo{person}{Senjuti~Basu Roy},
  \bibinfo{person}{Martine~De Cock}, {and} \bibinfo{person}{Ankur Teredesai}.}
  \bibinfo{year}{2016}\natexlab{}.
\newblock \showarticletitle{Predicting 30-Day Risk and Cost of "All-Cause"
  Hospital Readmissions}. In \bibinfo{booktitle}{{\em AAAI Workshop: Expanding
  the Boundaries of Health Informatics Using AI}}.
\newblock


\bibitem[\protect\citeauthoryear{Takahashi, Kumamaru, Jenkins, Saitoh,
  Morisawa, and Matsuda}{Takahashi et~al\mbox{.}}{2015}]%
        {TAKAHASHI2015286}
\bibfield{author}{\bibinfo{person}{Tetsuya Takahashi}, \bibinfo{person}{Megumi
  Kumamaru}, \bibinfo{person}{Sue Jenkins}, \bibinfo{person}{Masakazu Saitoh},
  \bibinfo{person}{Tomoyuki Morisawa}, {and} \bibinfo{person}{Hikaru Matsuda}.}
  \bibinfo{year}{2015}\natexlab{}.
\newblock \showarticletitle{In-patient step count predicts re-hospitalization
  after cardiac surgery}.
\newblock \bibinfo{journal}{{\em Journal of Cardiology\/}}
  \bibinfo{volume}{66}, \bibinfo{number}{4} (\bibinfo{year}{2015}),
  \bibinfo{pages}{286 -- 291}.
\newblock
\showISSN{0914-5087}
\showDOI{%
\url{https://doi.org/10.1016/j.jjcc.2015.01.006}}


\bibitem[\protect\citeauthoryear{van Walraven, Dhalla, Bell, Etchells, Stiell,
  Zarnke, Austin, and Forster}{van Walraven et~al\mbox{.}}{2010}]%
        {Walraven551}
\bibfield{author}{\bibinfo{person}{Carl van Walraven},
  \bibinfo{person}{Irfan~A. Dhalla}, \bibinfo{person}{Chaim Bell},
  \bibinfo{person}{Edward Etchells}, \bibinfo{person}{Ian~G. Stiell},
  \bibinfo{person}{Kelly Zarnke}, \bibinfo{person}{Peter~C. Austin}, {and}
  \bibinfo{person}{Alan~J. Forster}.} \bibinfo{year}{2010}\natexlab{}.
\newblock \showarticletitle{Derivation and validation of an index to predict
  early death or unplanned readmission after discharge from hospital to the
  community}.
\newblock \bibinfo{journal}{{\em Canadian Medical Association Journal\/}}
  \bibinfo{volume}{182}, \bibinfo{number}{6} (\bibinfo{year}{2010}),
  \bibinfo{pages}{551--557}.
\newblock
\showISSN{0820-3946}
\showDOI{%
\url{https://doi.org/10.1503/cmaj.091117}}
\showeprint{http://www.cmaj.ca/content/182/6/551.full.pdf}


\bibitem[\protect\citeauthoryear{Vedomske, Brown, and Harrison}{Vedomske
  et~al\mbox{.}}{2013}]%
        {Vedomske13}
\bibfield{author}{\bibinfo{person}{M.~A. Vedomske}, \bibinfo{person}{D.~E.
  Brown}, {and} \bibinfo{person}{J.~H. Harrison}.}
  \bibinfo{year}{2013}\natexlab{}.
\newblock \showarticletitle{Random Forests on Ubiquitous Data for Heart Failure
  30-Day Readmissions Prediction}. In \bibinfo{booktitle}{{\em 2013 12th
  International Conference on Machine Learning and Applications}},
  Vol.~\bibinfo{volume}{2}. \bibinfo{pages}{415--421}.
\newblock
\showDOI{%
\url{https://doi.org/10.1109/ICMLA.2013.158}}


\bibitem[\protect\citeauthoryear{Vooijs, Alpay, Snoeck-Stroband, Beerthuizen,
  Siemonsma, Abbink, Sont, and R{\"o}vekamp}{Vooijs et~al\mbox{.}}{2014}]%
        {Vooijs14}
\bibfield{author}{\bibinfo{person}{Martijn Vooijs},
  \bibinfo{person}{L.~Laurence Alpay}, \bibinfo{person}{B.~Jiska
  Snoeck-Stroband}, \bibinfo{person}{Thijs Beerthuizen},
  \bibinfo{person}{C.~Petra Siemonsma}, \bibinfo{person}{J.~Jannie Abbink},
  \bibinfo{person}{K.~Jacob Sont}, {and} \bibinfo{person}{A.~Ton
  R{\"o}vekamp}.} \bibinfo{year}{2014}\natexlab{}.
\newblock \showarticletitle{Validity and Usability of Low-Cost Accelerometers
  for Internet-Based Self-Monitoring of Physical Activity in Patients With
  Chronic Obstructive Pulmonary Disease}.
\newblock \bibinfo{journal}{{\em Interact J Med Res\/}} \bibinfo{volume}{3},
  \bibinfo{number}{4} (\bibinfo{date}{27 Oct} \bibinfo{year}{2014}),
  \bibinfo{pages}{e14}.
\newblock
\showDOI{%
\url{https://doi.org/10.2196/ijmr.3056}}


\bibitem[\protect\citeauthoryear{Wang, Cui, Chen, Avidan, Abdallah, and
  Kronzer}{Wang et~al\mbox{.}}{2017}]%
        {Wang17}
\bibfield{author}{\bibinfo{person}{Haishuai Wang}, \bibinfo{person}{Zhicheng
  Cui}, \bibinfo{person}{Yixin Chen}, \bibinfo{person}{Michael Avidan},
  \bibinfo{person}{Arbi~Ben Abdallah}, {and} \bibinfo{person}{Alexander
  Kronzer}.} \bibinfo{year}{2017}\natexlab{}.
\newblock \showarticletitle{Cost-sensitive Deep Learning for Early Readmission
  Prediction at A Major Hospital}.
\newblock


\bibitem[\protect\citeauthoryear{Wang, Robinson, Johnson, Zenarosa, Jayswal,
  Keithley, and Delaney}{Wang et~al\mbox{.}}{2014}]%
        {Wang2014}
\bibfield{author}{\bibinfo{person}{Hao Wang}, \bibinfo{person}{Richard~D.
  Robinson}, \bibinfo{person}{Carlos Johnson}, \bibinfo{person}{Nestor~R.
  Zenarosa}, \bibinfo{person}{Rani~D. Jayswal}, \bibinfo{person}{Joshua
  Keithley}, {and} \bibinfo{person}{Kathleen~A. Delaney}.}
  \bibinfo{year}{2014}\natexlab{}.
\newblock \showarticletitle{Using the LACE index to predict hospital
  readmissions in congestive heart failure patients}.
\newblock \bibinfo{journal}{{\em BMC Cardiovascular Disorders\/}}
  \bibinfo{volume}{14}, \bibinfo{number}{1} (\bibinfo{date}{07 Aug}
  \bibinfo{year}{2014}), \bibinfo{pages}{97}.
\newblock
\showISSN{1471-2261}
\showDOI{%
\url{https://doi.org/10.1186/1471-2261-14-97}}


\bibitem[\protect\citeauthoryear{Xu, Crammer, and Schuurmans}{Xu
  et~al\mbox{.}}{2006}]%
        {Xu2006}
\bibfield{author}{\bibinfo{person}{Linli Xu}, \bibinfo{person}{Koby Crammer},
  {and} \bibinfo{person}{Dale Schuurmans}.} \bibinfo{year}{2006}\natexlab{}.
\newblock \showarticletitle{Robust Support Vector Machine Training via Convex
  Outlier Ablation}. In \bibinfo{booktitle}{{\em Proceedings of the 21st
  National Conference on Artificial Intelligence - Volume 1}} {\em
  (\bibinfo{series}{AAAI'06})}. \bibinfo{publisher}{AAAI Press},
  \bibinfo{pages}{536--542}.
\newblock
\showISBNx{978-1-57735-281-5}
\showURL{%
\url{http://dl.acm.org/citation.cfm?id=1597538.1597625}}


\bibitem[\protect\citeauthoryear{Zhang}{Zhang}{2009}]%
        {Zhang08}
\bibfield{author}{\bibinfo{person}{Tong Zhang}.}
  \bibinfo{year}{2009}\natexlab{}.
\newblock \showarticletitle{Multi-stage Convex Relaxation for Learning with
  Sparse Regularization}.
\newblock In \bibinfo{booktitle}{{\em Advances in Neural Information Processing
  Systems 21}}, \bibfield{editor}{\bibinfo{person}{D.~Koller},
  \bibinfo{person}{D.~Schuurmans}, \bibinfo{person}{Y.~Bengio}, {and}
  \bibinfo{person}{L.~Bottou}} (Eds.). \bibinfo{publisher}{Curran Associates,
  Inc.}, \bibinfo{pages}{1929--1936}.
\newblock
\showURL{%
\url{http://papers.nips.cc/paper/3526-multi-stage-convex-relaxation-for-learning-with-sparse-regularization.pdf}}


\bibitem[\protect\citeauthoryear{Zheng, Zhang, Yoon, Lam, Khasawneh, and
  Poranki}{Zheng et~al\mbox{.}}{2015}]%
        {ZHENG20157110}
\bibfield{author}{\bibinfo{person}{Bichen Zheng}, \bibinfo{person}{Jinghe
  Zhang}, \bibinfo{person}{Sang~Won Yoon}, \bibinfo{person}{Sarah~S. Lam},
  \bibinfo{person}{Mohammad Khasawneh}, {and} \bibinfo{person}{Srikanth
  Poranki}.} \bibinfo{year}{2015}\natexlab{}.
\newblock \showarticletitle{Predictive modeling of hospital readmissions using
  metaheuristics and data mining}.
\newblock \bibinfo{journal}{{\em Expert Systems with Applications\/}}
  \bibinfo{volume}{42}, \bibinfo{number}{20} (\bibinfo{year}{2015}),
  \bibinfo{pages}{7110 -- 7120}.
\newblock
\showISSN{0957-4174}
\showDOI{%
\url{https://doi.org/10.1016/j.eswa.2015.04.066}}


\bibitem[\protect\citeauthoryear{Zhou, Shen, and Ye}{Zhou
  et~al\mbox{.}}{2011}]%
        {Zhou2011}
\bibfield{author}{\bibinfo{person}{Xi-chuan Zhou}, \bibinfo{person}{Hai-bin
  Shen}, {and} \bibinfo{person}{Jie-ping Ye}.} \bibinfo{year}{2011}\natexlab{}.
\newblock \showarticletitle{Integrating outlier filtering in large margin
  training}.
\newblock \bibinfo{journal}{{\em Journal of Zhejiang University SCIENCE C\/}}
  \bibinfo{volume}{12}, \bibinfo{number}{5} (\bibinfo{date}{04 May}
  \bibinfo{year}{2011}), \bibinfo{pages}{362}.
\newblock
\showISSN{1869-196X}
\showDOI{%
\url{https://doi.org/10.1631/jzus.C1000361}}


\bibitem[\protect\citeauthoryear{Zolfaghar, Meadem, Teredesai, Roy, Chin, and
  Muckian}{Zolfaghar et~al\mbox{.}}{2013}]%
        {Zolfaghar13}
\bibfield{author}{\bibinfo{person}{K. Zolfaghar}, \bibinfo{person}{N. Meadem},
  \bibinfo{person}{A. Teredesai}, \bibinfo{person}{S.~B. Roy},
  \bibinfo{person}{S.~C. Chin}, {and} \bibinfo{person}{B. Muckian}.}
  \bibinfo{year}{2013}\natexlab{}.
\newblock \showarticletitle{Big data solutions for predicting
  risk-of-readmission for congestive heart failure patients}. In
  \bibinfo{booktitle}{{\em 2013 IEEE International Conference on Big Data}}.
  \bibinfo{pages}{64--71}.
\newblock
\showDOI{%
\url{https://doi.org/10.1109/BigData.2013.6691760}}


\end{thebibliography}
